\documentclass[rmp,aps,groupedaddress,preprint,floatfix]{revtex4}
\usepackage{graphicx}

\begin{document}

\title{The nature of ferroelectricity under pressure}
\author{Igor A. Kornev}
 \altaffiliation[Also at ]{Novgorod State University, Russia}
\email{ikornev@uark.edu}
\author{Laurent Bellaiche}%
\affiliation{Physics Department, University of Arkansas,
Fayetteville, Arkansas 72701, USA}%

\date{\today}

\begin{abstract}

Advances in first-principles computational approaches have, over
the past fifteen years, made possible the investigation of
physical properties of ferroelectric systems. In particular, such
approaches have led to a \emph{microscopic} understanding of the
occurrence of ferroelectricity in perovskite oxides at ambient
pressure. In this paper, we report {\it ab-initio} studies on the
effect of hydrostatic pressure on the ferroelectricity in
perovskites and related materials. We found that, unlike commonly
believed, these materials exhibit ferroelectricity at high enough
pressure. We also  explained in details the (unusual) nature of
this ferroelectricity.
\end{abstract}
\maketitle


\narrowtext

\marginparwidth 2.7in
\marginparsep 0.5in

\section{Introduction.}

ABO$_{3}$ perovskites form one of the most important classes of
materials because they can exhibit a broad range of properties,
e.g. superconductivity, magnetism, ferroelectricity,
piezoelectricity, dielectricity and multiferroism. Such properties
can be varied -- and thus optimized to generate various devices
with great performance and diverse functionalities -- thanks to
many factors. Examples of such factors are external magnetic and
electrical fields, atomic substitution, chemical ordering
and pressure.\\

In a classic paper~\cite{samara:1767}, Samara \textit{at al}. were
able to explain the effect of hydrostatic pressure on displacive
phase transitions associated with soft zone-center transverse
optic (TO) and zone-boundary phonons. Among the most striking
pressure effects is the decrease of the transition temperature
with pressure and ultimate vanishing of
ferroelectricity~\cite{samara:1767}. As it was shown, the
vanishing of ferroelectricity with pressure can be readily
understood from the well-known soft-mode
theory~\cite{ginzburg:1037} where the soft mode frequency depends
on the short-range and Coulomb interactions. In this case, the
vanishing of ferroelectricity is associated with a much more rapid
increase of the short-range interactions than the long-range
interactions with increasing pressure. As a result, the harmonic
soft-mode frequency becomes less negative with increasing
pressure.  It is interesting to note that for the zone boundary
modes the roles of the short-range and long-range forces are
reversed, which leads to an enhancement of these modes when
increasing pressure. Despite its simplicity, this model could
successfully explain high pressure behavior of many
compounds~\cite{samara:1767, sani:020105}. [Note that
``long-range'' interactions denote interactions that persist even
at distances where the electronic wavefunctions no longer overlap.
Such long-range interactions have a distance dependency that is
inversely proportional to some power of the interionic distance.
``Short-range'' interactions only occur when the electronic
wavefunctions overlap. They typically have a distance dependency
that follows an exponential form at large distances, and are
inversely proportional to  some power of the interionic distance
at smaller distances.

The microscopic understanding of the occurrence of
ferroelectricity in perovskite oxides has been substantially
improved by first-principles electronic band-structure techniques.
In particular, Cohen \& Krakauer~\cite{cohen:6416,cohen:136,
cohen:65} showed that a delicate balance exists between Coulomb
interactions that favor ferroelectric distortions and the
short-range repulsions that favor the undistorted high-symmetry
structure. It it has been established that the balance between
different competing interactions can be tipped by relatively small
effects that are related to covalency, namely hybridization
between $p$ orbitals of the oxygen atoms and the  $d$ orbitals of
the B atoms. In fact, such covalency is  essential for stabilizing
the ferroelectric distortion or, equivalently, triggers the
soft-mode instability. The reason for this stabilization is that
the hybridization weakens the short-range repulsions. This fact
has been elegantly demonstrated by Posternak \textit{et
al.}~\cite{posternak:8911} -- who showed that the ferroelectric
instability disappears in KNbO$_3$ when the interaction between
the O $2p$ and Nb $4d$ states is artificially suppressed. Since
then, similar conclusions have been drawn for other ferroelectric
perovskites. Moreover, an additional Pb-O hybridization is the key
factor to explain why PbTiO$_3$ has deeper ferroelectric
double-well potentials  than BaTiO$_3$~\cite{cohen:65,
kuroiwa:217601}.  The important role of covalent bonding is also
clearly reflected in the anomalously large values of the Born
dynamical charges~\cite{PhysRevLett.72.3618, ghosez:6224}.

In this paper, we use first principles calculations to investigate
many insulating perovskites and related materials under
hydrostatic pressure, and show that the conclusion about the role
of the different interactions mentioned above is indeed well
justified at low pressure but is no longer valid at high
pressure~\cite{kornev:196804}. More precisely, our study reveals
(1) the \emph{absence} of a cubic paraelectric phase in favor of a
polar phase under high-enough hydrostatic pressures, and (2) the
underlying physical mechanisms responsible for this overlooked
phenomenon.

\section{Computational Details.}

The \textit{ab initio} calculations are performed in the framework
of the density functional theory, mostly within the local-density
approximation (LDA) but also sometimes within the gradient
generalized approximation (GGA)~\cite{Perdew:8800,
PhysRevLett.77.3865}. We have employed the pseudopotential method
using the ABINIT~\cite{Gonze:478}, CUSP ~\cite{PhysRevB.47.1651,
PhysRevB.61.7877}, and PWscf~\cite{pwscf} packages. The
calculations have been done with the extended norm-conserving
pseudopotentials of Teter~\cite{teter:5031} or with the
Vanderbilt-type ultrasoft pseudopotentials \cite{vanderbilt:7892}
-- always incorporating the semi-core states in the valence,
unless specified otherwise. The wavefunction has been expanded in
planewaves up to a kinetic energy cutoff of 120 Rydberg when using
the extended norm conserving pseudopotentials and of 40 Rydberg
when using the ultrasoft pseudopotentials. We  use the
Ceperley-Alder exchange and correlation~\cite{ Ceperley:566} as
parameterized by Perdew and Zunger~\cite{Perdew:5048} or the
Teter-Pade parametrization ~\cite{Goedecker:1703, Perdew:13244}.
For the ground state calculations, integrals over the simple cubic
Brillouin zone have been replaced by sums on a 6 x 6 x 6
Monkhorst-Pack special k-point mesh~\cite{PhysRevB.13.5188}. We
determined: (i) the total energy for different volumes of the unit
cell allowing for full relaxation of the cell shape and the
positions of the atoms inside the unit cell, in order to determine
the evolution of some phases as a function of hydrostatic
pressure, (ii) the phonon spectrum of specific structures by using
both finite-difference (frozen-phonon) and linear-response
approaches~\cite{Gonze:3603}, (iii) the Born effective charges,
the dielectric constants, and (iv) the electronic band-structure
and density of states. It should be noted that we checked that our
results are qualitatively independent on technical details such as
the used exchange-correlation functional, pseudopotential types,
kinetic energy cutoff or softwares.

\section{Results and discussion}
\subsection{PbTiO$_3$ under pressure}

Lead titanate PbTiO$_{3}$ is one of the most important and best
studied ferroelectric materials adopting the simple-cubic
perovskite structure at high temperature. It is known that
PbTiO$_3$ undergoes a single ferroelectric transition from the
cubic to a tetragonal phase below 766 K. In fact, it appears to be
a textbook example of a displacive ferroelectric transition~\cite{
burns:3088,lines:1977}.

Hydrostatic pressure \textit{P} reduces, and even annihilates for
high enough value, ferroelectricity (FE) in
perovskites~\cite{sanjurjo:7260, sani:020105}, \textit{according
to the conventional theory}. Such annihilation seems consistent
with experimental X-ray diffraction studies
~\cite{zha:3705,sani:10601} and first-principles
calculations~\cite{wu:037601} that reported a ferroelectric to
paraelectric phase transition in lead titanate under a
``moderate'' (i.e. around 10-20 GPa) pressure. On the other hand,
one of our recent studies~\cite{kornev:196804} -- that combines
experimental techniques and \textit{ab-initio} simulations --
revealed that perovskites enhance their ferroelectricity  as
pressure increases {\it above a critical value} (which is around
30 GPa in PbTiO$3$).

More specifically, we performed 0 K calculations on PbTiO$_3$
within LDA and for a wide pressure range. The calculated
pressure-volume data were fitted to a Birch equation of
state~\cite{Birch:4949}. Here, we present the results for 7
phases: the {\it paraelectric} cubic $Pm\overline{3}m$ state; the
{\it ferroelectric}, rhombohedral $R3m$, and tetragonal $P4mm$
phases; the {\it antiferrodistortive} tetragonal $I4/mcm$ and
rhombohedral $R\overline{3}c$  phases; and the tetragonal $I4cm$
and rhombohedral $R3c$ phases that can exhibit {\it both} polar
\emph{and} antiferrodistortive degrees of freedom.

Figure~\ref{fig:Fig1}\textit{a} shows the predicted pressure
dependency of the difference
\textit{\ensuremath{\Delta}}\textit{H} = \textit{H} --
\textit{H}$_{\mathit{cubic}}$ between the
\textit{H}$_{\mathit{cubic}}$ enthalpy of the paraelectric cubic
state and the \textit{H} enthalpies of the other 6 phases. For the
lowest (\textit{P}) pressures, only ferroelectric distortions
exist in $I4cm$ and $R3c$, which are thus identical to $P4mm$ and
$R3m$, respectively. The $P4mm$ phase has the lowest
\textit{\ensuremath{\Delta}}\textit{H}, as consistent with the
fact that it is the well-known ground state of PbTiO$_{3}$.
Figs.~\ref{fig:Fig1}(\textit{b, c}) show that, as expected in
perovskites~\cite{samara:1767, sani:020105}, significant
antiferrodistortive deformations occur in the $I4cm$ and $R3c$
phases when increasing \textit{P} above \ensuremath{\sim}3 GPa.
The reason leading to the TiO$_6$ octahedra rotation is clear: due
to the TiO$_6$ octahedra rotation, the Ti-O bonds perpendicular to
the rotational axis can relax towards their equilibrium lengths,
which minimizes the bond compression. In other words, the rotation
relaxes the compression of the Ti-O bond perpendicular to the
rotational axis. On the other hand, Fig.~\ref{fig:Fig1} also
indicates an unexpected result, namely that \textit{increasing
pressure above a critical value enhances ferroelectricity}!
Indeed, an increase in \textit{P} from 0 to \ensuremath{\sim}20
GPa leads to a \textit{\ensuremath{\Delta}}\textit{H} becoming
less negative in the polar phases - as consistent with the
commonly-believed pressure-induced reduction of
FE~\cite{samara:1767, sanjurjo:7260, sani:020105}, but the
\textit{\ensuremath{\Delta}}\textit{H --vs.-P} curve of these
phases adopts an opposite behavior above
\textit{P}$_{\mathit{c}}$\ensuremath{\sim}30 GPa - revealing that
PbTiO$_{3}$ wants to become more and more ferroelectric after this
critical pressure. We found that rhombohedral states are the most
stable ones at high pressure~\cite{kornev:196804} and at 0K. The
non-monotonic behavior of FE versus pressure  is
 emphasized by the
pressure dependency of the calculated \textit{c/a} axial ratio in
P4mm (see Fig.~\ref{fig:FigCA}):  this ratio first strongly decreases with
\textit{P} until reaching a cubic-like value \ensuremath{\sim}1 in
the vicinity of \textit{P}$_{\mathit{c}}$, and then it gradually
increases as \textit{P} further increases. Our predictions are thus in
sharp contrast to the common expectation that the application of
hydrostatic pressure favors a cubic paraelectric crystal structure.

\subsection{Possible  driving mechanism(s) for the high-pressure ferroelectric instability in PbTiO$_3$.}

In order to identify the driving mechanism(s) of the ferroelectric
instability under high pressure, we decided to investigate in more
details (1) the cubic paraelectric $Pm\overline{3}m$ phase and (2)
the tetragonal ferroelectric $P4mm$ phase.
Fig.~\ref{fig:Fig1}\textit{b} shows that the magnitude of the
polarization above the turning point is smaller than that below
$\simeq$ 10GPa. Based on this fact, one {\it might} assume that
the polarization is no longer the primary order parameter at high
pressure or, equivalently, that the ferroelectric instability is
driven by the coupling with other degrees of freedom, such as  (i)
the antiferrodistortive and (ii) elastic variables.

Possibility (i) is a primary suspect because antiferrodistortive
degrees of freedom are known to increase with pressure in
ferroelectric perovskites~\cite{sani:020105}. In this case, the
ferroelectric instability could appear as a result of a specific
interaction between the antiferrodistortive and ferroelectric
distortions. However, these two degrees of freedom usually {\it
compete} in perovskites~\cite{6384307}. Furthermore, as shown in
Figure~\ref{fig:Fig1}, the ferroelectric instability exists even
without taking into account the antiferrodistortive degrees of
freedom. For instance, after the turning point around $\simeq$
30GPa, the $P4mm$ phase (that does not contain any rotation of the
oxygen octahedra) is more stable than the cubic phase. These facts
thus rule out the antiferrodistortive degrees of freedom to be the
driven force for the ferroelectric instability.

Possibility (ii) is also a candidate for explaining high-pressure
FE because perovskite ferroelectrics are well-known to exhibit a
coupling between polarization and strain. An elastic {\it
instability} may thus lead to the occurrence of a nonzero
polarization. To check such hypothesis, we performed \textit{ab
initio} calculations of the elastic tensor in the cubic phase
using the ABINIT package and found \emph{no elastic instability}
in the entire pressure range. Fig.~\ref{fig:Elast} shows the
pressure dependence of the elastic stiffness coefficients $c_{11}$
, $c_{12}$ , and $c_{44}$. The $c_{11}$ and $c_{12}$ coefficients
increase linearly with increasing pressure. On the other hand, the
$c_{44}$ coefficient starts to decrease gradually above the
critical pressure but remains positive at all investigated
pressures (that is, up to $\simeq$ 150 GPa) . Furthermore, these
elastic stiffness coefficients satisfy the generalized elastic
{\it stability} criteria for cubic crystals under hydrostatic
pressure [see, for example, Ref. \cite{0953-8984-9-41-005}]:
$c_{11}+2c_{12}>0$, $c_{11}-c_{12}>0$, $c_{44}>0$. Moreover, we
found that the $P4mm$ phase \emph{without} lattice relaxation is
more stable at high pressure than the cubic phase. Therefore, we
may now assume that the polarization {\it is} the primary order
parameter even after the turning point. Elucidating the mechanisms
of the high-pressure ferroelectric instability should thus involve
the consideration of the soft transverse optical mode alone. (Note
that we can also conclude that the electromechanical coupling that
become enormous in the transition regions from the tetragonal to
high pressure phases of PbTiO$_3$~\cite{aps2006:wu} is not
responsible for the ferroelectric instability).

\subsection{Lattice Dynamical Properties of PbTiO$_3$ and of Related Perovskites.}

It is well-known that the Born effective charges are of primary
interest for the soft TO mode behavior when a competition between
the Coulomb and non-classical short-range forces is important. In
the following, we will denote O$_\parallel$ the oxygen atoms
located between two B-atoms along the z-direction and  O$_\perp$
the oxygen atoms located between two B-atoms in the perpendicular
directions -- when slightly displacing the atoms along the z-axis
when computing the Born effective charges. Our calculations of the
Born effective charges as a function of pressure in the cubic
phase (see Fig.~\ref{fig:ZB}) show that the anomalously large
values of Ti and O$_\parallel$ persist throughout the pressure
range of interest, although these effective charges decrease  with
increasing pressure {For instance, their values are reduced by
$\sim1.4\%$ at $\sim25$ GPa with respect to their values at 0GPa].
On the other hand, the effective charges of Pb and O$_\perp$ ions
increase with pressure, and deviate from their values at zero
pressure by $\sim2\%$ at $\sim25$ GPa. Interestingly, it was
demonstrated by Ghosez \textit{et al.}~\cite{Ghosez:713} that a
change corresponding to a reduction of the order of 1\%  of the Ti
effective charges is enough to suppress the ferroelectric
instability in BaTiO$_3$. From this point of view, the tendency of
the PbTiO$_3$ system to become cubic as the pressure first
increases from zero is quite understandable. However, since after
the turning point the Ti and O$_\parallel$ effective charges
continue to decrease with pressure, one might expect that the
system would continue to stay cubic {\it which is definitely not
the case}. Therefore, the pressure behavior of the effective
charges or, equivalently, the strengths of the Coulomb interaction
{\it alone} is not sufficient to understand high-pressure FE.

We now turn to Fig.~\ref{fig:Omega}\textit{a} that displays the
LDA-simulated pressure behavior of the zone-center transverse
optical (TO) mode frequency  in the cubic $Pm\overline{3}m$ phase
of PbTiO$_{3}$. The square of such frequency is always negative,
which indicates that PbTiO$_{3}$ has a ferroelectric instability
and thus can not be cubic for any pressure at 0K. More
importantly, the magnitude of this square first decreases with
\textit{P} and then increases above the critical pressure
(\textit{P}$_{\mathit{c}}$) - confirming that PbTiO$_{3}$ has to
become more and more ferroelectric above
\textit{P}$_{\mathit{c}}$. To understand further the high-pressure
ferroelectric instability it is advantageous to divide the
dynamical matrix into (1) the Coulomb part (following the
formalism for the rigid ion model ) that can easily be found once
the Born effective charges and dielectric tensor of the system
have been calculated~\cite{giannozzi:7231}, and (2) the
(remaining) second part -- which is deduced by subtracting the
first part from the total dynamical matrix. This second part thus
differentiates the real material from its rigid ions picture, and
gathers the (non-classical) short-range interactions.
Figure~\ref{fig:Omega}\textit{b} shows the calculated contribution
of these two parts to the zone-center soft TO phonon frequency in
the cubic phase of PbTiO$_{3}$ and BaTiO$_{3}$, as a function of
\textit{P}. Below \textit{P}$_{\mathit{c}}$ and as commonly
expected~\cite{samara:1767}, the Coulomb interactions favor FE
while short-range interactions tend to annihilate it - since they
have a negative and positive frequency square, respectively - and
the decrease of FE when increasing \textit{P} is driven by the
cost of short-range interactions becoming predominant over the
gain associated with long-range interactions.
Figure~\ref{fig:Omega}\textit{b} also reveals that FE above
\textit{P}$_{\mathit{c}}$ dramatically differs from the
low-pressure ``normal'' ferroelectricity because of the
\emph{reverse} sign of Coulomb and short-range parts: it is now
short-range effects (which are electronic in nature) that favor FE
at high pressure while Coulomb interactions (that involve ions
``dressed by electrons'') prefer paraelectricity!

Figure~\ref{fig:Omega}\textit{b} indicates that the results that
we present are qualitatively general for perovskite ferroelectrics
and not just an artifact of studying lead-contained perovskites,
since they also apply to BaTiO$_{3}$. This fact is further
emphasized in Figures~\ref{fig:Omega3} that report our 0\,K
\emph{ab-initio} predictions for the eigenvalue $\lambda$ and
eigenvectors ($\xi_A$, $\xi_B$, $\xi_{O_\perp}$,
$\xi_{O_{\perp}}$, $\xi_{O_{\parallel}}$) of the second-derivative
matrix defined in Ref. \cite{King-Smith:5828} and associated with
the soft-mode  in cubic PbTiO$_3$,  BaTiO$_3$ and BaZrO$_3$, as a
function of pressure. (It is important to realize that the last
two materials are {\it lead-free}). A {\it negative} $\lambda$
corresponds to a ferroelectric instability while the magnitude of
the eigenvectors quantifies the contribution of the different ions
to FE, if any. All the data displayed in Figs.~\ref{fig:Omega3}
are obtained by using the local density approximation (LDA) except
in parts (a) and (d) for which we also report results from the use
of the generalized gradiant approximation (GGA).
Figs.~\ref{fig:Omega3} undoubtedly show that {\it all} studied
materials should be ferroelectric at high enough pressure, thus
confirming that one has to rule out a possible suggestion that
pressure-induced FE only occurs in Pb-based materials. One can
further notice that $\xi_A$ has a very small magnitude at high
pressure in PbTiO$_3$, BaTiO$_3$ and BaZrO$_3$. This fact clearly
implies that the $A$ atoms contribute very little to high-pressure
ferroelectricity. Our Figs.~\ref{fig:Omega3} also nicely summarize
the different behaviors one can get as pressure increases in
transition-metal perovskites, as a result of our discovered
non-monotonic pressure behavior: the existence of FE at any
pressure (PbTiO$_3$, see LDA calculation of
Fig.~\ref{fig:Omega3}(a)), the disappearance and then reentrance
of FE under pressure (BaTiO$_3$, see Fig ~\ref{fig:Omega3}(b)),
and the occurrence of FE at high pressure in a nominally
paraelectric compound (BaZrO$_3$, see Fig.~\ref{fig:Omega3}(c)).
Moreover, Figures~\ref{fig:Omega3}(d-f) clearly shows that the
eigenvectors of $B$ and $O$ atoms have a large (and mostly
material-independent!) magnitude at high pressure, which  points
to an ``universal'' microscopic explanation for high-pressure FE.
Finally, such pressure-induced phenomenon is also found when
performing \textit{ab-initio} calculations in KNbO$_3$, NaNbO$_3$,
SrTiO$_3$, CaTiO$_3$, as shown in Fig.~\ref{fig:OmegaAll},
Pb(Zr,Ti)O$_3$ and some Bi-based perovskites (not shown here). Not
also that the fact that the A atom contributes little to
high-pressure FE is clearly confirmed in
Fig.~\ref{fig:OmegaWO}--that shows that A-free WO$_3$ material
also enhances its FE at high enough pressure.

\subsection{Electronic Structures under Pressure}

We now resume our ``detective'' work and look for the microscopic
origin of the observed high-pressure FE. As it was shown above,
the instability is related to the non-classical short-range
interactions. As a result, we will now mostly concentrate on {\it
electronic} properties.

\subsubsection{Influence of semicore states.}

We first investigate the role played by the semicore states via
Fig.~\ref{fig:FigSCr} that shows the predicted pressure behavior
of phonons in PbTiO$_3$ in the cubic phase {\it when neglecting to
incorporate the semicore states of Pb and Ti as valence
electrons}. Technically, the \textit{ab-initio} simulations
without the semicore states have been performed with the
Troullier-Martins norm-conserving pseudopotentials within ABINIT.
Fig.~\ref{fig:FigSCr} indicates that the semicore states Pb
$5d^{10}$ and Ti $3s^23p^6$ (1) must be included explicitly to
produce the well-known ferroelectric instability at zero pressure
-- as consistent with previous first-principles
works~\cite{cohen:65, King-Smith:5828} but (2) are {\it not} the
reason for  the existence of a turning point and of high-pressure
FE  (Note, however, that the magnitude of the instability is of
course not the same when neglecting or incorporating  these
semicore states in the valence electrons). Note that the fact that
Pb $5d^{10}$ orbitals are not responsible for the existence of the
ferroelectric instability  at high pressure could also have been
guessed thanks to Fig.~\ref{fig:OmegaWO} that shows that tungsten
trioxyde WO$_3$ (which is a defect-perovskite structure having
{\it no} $A$-atom)  also exhibits a highy-pressure FE. In fact,
this latter figure, as well as
Figs.~(\ref{fig:Omega3},\ref{fig:OmegaAll}), indicates that the
electrons of interest for high-pressure FE should belong to the O
and/or B ions.

\subsubsection{Influence of the $d$-states of $B$ atoms.}

Figure~\ref{fig:FigNOd} reports the LDA-predicted pressure
behavior of the soft-mode eigenvalues of the second-derivative
matrix in cubic bismuth aluminate BiAlO$_3$ and magnesium silicate
MgSiO$_3$. Note that both materials, unlike the ones considered
above, do not possess any B-atom having \emph{occupied} $d$
electrons. Interestingly, no turning point is observed in
MgSiO$_3$ - under pressure up to 300 GPa. As a matter of fact, we
found that the TO soft mode at the $\Gamma$ point linearly
increases with pressure in this material and become positives at
around 30 GPa - which is consistent with previous studies of
MgSiO$_3$~\cite{parlinski:49}. The same qualitative behavior is
found in BiAlO$_3$ under high pressure up to 900 GPa. We thus
arrive at the conclusion that the $d$ states of the B atoms are of
primary importance for understanding the physics behind the
high-pressure ferroelectric instability.

\subsubsection{ Electronic band structure and density of states under pressure}

To further elucidate the nature of the high-pressure ferroelectric
instability and the role played by the electronic structure, we
have calculated the total and atomic site projected densities of
states (PDOS) of ideal cubic perovskites. Figs.~\ref{fig:PTOdos}
display some partial electronic density of states for cubic
PbTiO$_{3}$ under different pressures. Note that the contribution
of Pb orbitals is not shown because, as discussed above, Pb was
found to be relatively ferroelectric-inactive at high pressure.

Figures~\ref{fig:PTOdos}(\textit{a-c}) show the low-lying states
ranging between -22 and -15 eV  (that solely originate from the O
\textit{2s} orbitals in an ionic picture), as well as the topmost
group of valence bands (that exhibits the well-known hybridization
between O \textit{2p} and Ti \textit{3d} orbitals
~\cite{cohen:6416, cohen:136}). As it is known, the topmost states
correspond mainly to $\sigma$ (bonding) and $\pi$ (non-bonding)
orbitals. Moreover, the states with large Ti $3d$ contribution are
mainly concentrated in the lower (bonding) part of the topmost
region. This O \textit{2p} and Ti \textit{3d} hybridization
persists throughout the pressure range of interest. The integrated
PDOS of this group have a smooth and small pressure-dependency
thus ruling out that high-pressure FE is caused by this \textit{p
-- d} hybridization. At the same time the Ti \textit{3d} state
contribution to the low-lying O \textit{2s}-like bands has
increased considerably. Thus, as the pressure increases, the
admixture of Ti \textit{3d} states in the occupied bands seems to
evolve differently for the occupied low-lying and topmost bands.
More precisely, the evolution of O $2s$ - Ti $3d$ and O $2p$ - Ti
$3d$ mixing is nicely seen in the trends displayed by the
\emph{integrated} (over all the occupied valence bands) PDOS (to
be denoted by IPDOS), see Fig.~\ref{fig:ipdosPTO1}(\textit{a}).
The IPDOS of $O~2s$ and $Ti~3d$ vary in a correlated way with
pressure, the former decreasing and the latter increasing. On the
other hand, the IPDOS of $O~2p$ remains more or less constant,
apart from the highest pressures. These tendencies are even more
clearly seen in Fig.~\ref{fig:ipdosPTO1}(\textit{b}), where IPDOS
are normalized with respect to their corresponding values at zero
pressure. Furthermore, Fig.~\ref{fig:ipdosPTO1}(\textit{c})
reports the contributions to the IPDOS coming from the low-lying
states (located between -22 and -15 eV) and from the topmost group
of the valence bands. The most striking effect is that Ti
\textit{3d} orbitals (and more precisely their
\textit{e}$_{\mathit{g}}$ parts) dramatically increase (in a
nonlinear way) their mixing with the O \textit{2s} orbitals above
\textit{P}$_{\mathit{c}}$. The admixture of Ti $3d$ states in the
occupied levels of the electronic structure releases O $2s$ - O
$2s$ antibonding states above the Fermi level, thus optimizing the
band energy. A similar behavior of integrated PDOS was found in
BaTiO$_3$ under pressure, as reported in Fig.~\ref{fig:ipdosBTO1}.

A rapid relative increase of the $O$-$s$  and $B$-$d$ mixing can
easily be understood as follows. The lobes of the Ti  $3d-t_{2g}$
orbitals point between the oxygen ligands (whereas the Ti $3d-e_g$
orbitals point directly towards the ligands), forming the $\sigma$
bonding. Hence, the overlap with O $2s$ orbitals will be
significant at high enough pressure for the Ti  $3d-e_g$ states,
which results in local repulsion between overlapping charge
densities. From one hand, this repulsive interaction will cost
some energy. From the other hand, some of the repulsion is
compensated by hybridization in the resulting bonding states --
which in turn leads to considerable amount of O $2s$ state in the
conduction band. Taking into account the different interatomic
dependence of these two competing interactions, one might expect
that the need to avoid the large local repulsion will win at some
pressure, enabling atomic distortions  and thus giving rise to
ferroelectricity.

We would like to point out that the important role played by the
small amount of $2s$ admixture with the $e_g$ orbitals of
transition-metal cations surrounded by anions (like chlorine,
bromine, fluorine, or oxygen) was also previously recognized in
Ref.~\cite{moreno:14423}. Moreover, recent DFT calculations
~\cite{hussermann:1472} revealed the key role of $s - d$ mixing on
the stability of high-pressure structures in P, As, Sb, and Bi.

\subsection{Topological analysis of the electron density in PbTiO${_3}$.}

To get more insight into the chemical bondings of PbTiO${_3}$
under pressure, we consider the topological properties of the
electron density $\rho(r)$. The {\it total} density was restored
by adding atomic core electron density to the valence density
obtained in pseudopotential calculations. The analysis was
performed in terms of the atom-in-molecules Bader theory
~\cite{bader:1990} which is based upon the so-called critical
points, $r_{cp}$, of the electron density. At these points, the
gradient of the electronic density is null and  the Hessian matrix
$\partial^2\rho(r)/\partial x_i \partial x_j$ is characterized by
its three eigenvalues, $\lambda_k$. Interatomic interactions can
be adequately described by the topological properties of the
electron density at the so-called (3,-1) critical points (or bond
critical points) for which the three eigenvalues of the Hessian
matrix are non-zero and the algebraic sum of their signs is $-1$.
Bond critical points occur as saddle points along bond paths
between pairs of bonded atoms and can be used to define atomic
coordination. The characteristics of the bond critical points in
PbTiO$_3$ at 0 GPa are listed in Table 1.

\begin{table}
\caption{ Characteristics of some critical points (CP) in PbTiO$_3$ at
$a=7.35~a.u.$ (zero pressure);  $\nabla^2\rho(r_{cp}) = \lambda_1 + \lambda_2 +
\lambda_3$, where $\lambda_1$ and $\lambda_2$ are perpendicular to
the bond line curvatures of the electron density and $\lambda_3$
is parallel. Values are given in atomic units.} \label{tabI}
\begin{tabular}{c|cc|ccc|c}
\cline{1-7}

$r_{cp}$ & $\rho(r_{cp}$) &$\nabla^2\rho(r_{cp}$)&
$\lambda_1$ & $\lambda_2$ & $\lambda_3$  & Type of CP\\
\tableline
 $(0.234, 1/2, 1/2)$\tablenotemark[1]  & 0.132  & 0.590 & -0.219 & -0.219 & 1.028 & (3,-1)\\
 $(0, 0.267, 0.267)$\tablenotemark[2]  & 0.028  & 0.078 & -0.020 & -0.022 & 0.120 & (3,-1)\\

\end{tabular}
\tablenotetext[1]{Ti-O line.} \tablenotetext[2]{Pb-O line.}
\end{table}

As expected for cubic perovskite [see, for instance,
Ref.~\cite{zhurova:567}], we found six (3,-1) critical points on
the Ti-O lines and twelve (3,-1) critical points on the Pb-O lines
at any pressure. Interestingly, all the (3, -1) critical
points in PbTiO$_3$ are characterized by the positive Laplacian of
the electron density which, according to Bader~\cite{bader:1990},
is an indication of the closed-shell interaction between the
corresponding atoms.

Analyzing the topological characteristics available in the
literature, it was found ~\cite{zhurova:567} that the bonds,
typically identified as {\it ionic}, are characterized by the
electron density value at the (3,-1) critical point in the range
of  0.07--0.25 e~\AA$^{-3}$ and a specific range of the ratio of
the perpendicular and parallel curvatures of the electron density:
$0.12 < |\lambda_1|/\lambda_3 < 0.17$. According to Table I, the
Pb-O bond (at 0 GPa) fits well the first condition since
$\rho(r_{cp}) \approx 0.1908$ e~\AA$^{-3}$ and closely fits the
second one if we take $|(\lambda_1+\lambda_2)/2|/\lambda_3\approx
0.176 $ rather than $|\lambda_1|/\lambda_3$. Thus, the Pb-O
interaction (at zero pressure) can be considered as mostly ionic
with small covalent admixture. On the other hand, the value of
$\rho(r_{cp}) \approx 0.9$ e~\AA$^{-3}$ and the ratio
$|\lambda_1|/\lambda_3 = 0.213$ at the (3,-1) critical points on
the Ti-O line imply that the Ti-O bond is {\it not} ionic in
nature (at zero pressure).

In order to characterize the chemical bonds in more details (and
for different pressures), we employ the recently proposed
classification scheme of interatomic interactions
~\cite{espinosa:5529}. According to Bader~\cite{bader:1990} the
Laplacian function $\nabla^2\rho(r_{cp})$ is related to the
potential energy density $v(r_{cp})$ (of negative value) -- which
represents the capacity of the system to concentrate electrons at
the critical point -- and the kinetic energy density $g(r_{cp})$
(of positive value) -- which gives the tendency of the system to
dilute electrons at the same point~\cite{espinosa:170}  -- through
the local form of the virial theorem $1/4\nabla^2\rho(r_{cp}) =
2g(r_{cp})+v(r_{cp})$. Note that for pressure different from zero,
the virial theorem should include an additional term proportional
to this P pressure: in the cubic case, one should add the $+3P$
term to the lefthand side of this latter expression.

The kinetic energy can be calculated using the following
expression for closed-shell interactions at the bond critical
points ~\cite{abramov:264}: $g(r_{cp}) =
3/10(3\pi^2)^{2/3}\rho(r_{cp})^{5/3} +1/6\nabla^2\rho(r_{cp})$.
The potential energy density $v(r_{cp})$ is then calculated by
subtracting $g(r_{cp})$ from $1/4\nabla^2\rho(r_{cp})$. We also
computed the total electronic energy density $h(r)$ at the
$r_{cp}$ points as $h(r_{cp})=g(r_{cp})+v(r_{cp})$. Using the
adimensional $|v(r_{cp})|/g(r_{cp})$ ratio and the bond degree
$BD=h(r_{cp})/\rho(r_{cp})$, the atomic interactions can be
divided into three classes~\cite{espinosa:170}: (i) pure closed
shell for which $|v(r_{cp})|/g(r_{cp}) < 1$ (and for which the
 $\nabla^2\rho(r_{cp})>0$ and $h(r_{cp})>0$ inequalities stand), (ii) pure
shared shell for which $1 < |v(r_{cp})|/g(r_{cp}) < 2$ (and for
which $\nabla^2\rho(r_{cp})<0$ and $h(r_{cp})<0$) and (iii)
transit closed shell for which $|v(r_{cp})|/g(r_{cp}) > 2$ (with
$\nabla^2\rho(r_{cp})>0$ and $h(r_{cp})<0$).

Inside the ionic region (i),  the bond degree parameter is an
index of ionicity: the stronger the ionicity the greater the $BD$
magnitude. In regions (ii) and (iii), the $BD$ parameter measures
the covalency degree: the stronger the covalency the greater the
$BD$ magnitude. For instance, the ionic Li-F bond is characterized
by $|v(r_{cp})|/g(r_{cp})\approx0.76$ and $BD\approx+0.41$ while
the covalent C-O bond in CO(NH$_2$)$_2$ is such as
$|v(r_{cp})|/g(r_{cp})\approx 2.64$ and
$BD\approx-0.29$~\cite{tsirelson:632}. Fig.~\ref{fig:BD} reports
the bond degree parameter versus $|v(r_{cp})|/g(r_{cp})$ ratio (as
determined from our first-principles calculations) for the Ti-O
and Pb-O bonds. This figure confirms that the Pb-O bond lies close
to the pure closed shell region (i.e., it is mostly ionic) at
0GPa, and reveals that such bond becomes more ionic at higher
pressure. On the other hand, the Ti-O bond lies close to the
middle of a transit region (ii), implying that the Ti-O bond is
neither purely covalent nor purely ionic. Fig.~\ref{fig:BD} also
shows that the covalency degree of the Ti-O bond increases with
pressure, which is consistent with our band structure calculations
discussed above in subsection D.3. Interestingly, since this
covalency increases with pressure, there might be a possibility
for the phase transition to occur due to the drastic collapse of
the oxygen \textit{ionic} size (which is related to charge
transfer to oxygen). For example, the Shannon ionic radius of
O$^{2-}$ is 1.23 \AA~\cite{shannon:751}, whereas its covalent
radius is 0.73 \AA~\cite{sanderson:1962}. If the size of oxygens
were small compared to the allowed space then the oxygen atoms
would be weakly bounded in the lattice and a multi-well potential
could be expected [similar to PbTe:Ge~\cite{murase:725,
PhysRevB.23.2741} and PbTe$_{1-x}S_x$~\cite{abdullin:998}]. The
mechanism for the off-center instability of the O ion could  be
viewed in this case as an imbalance between the decreased (due to
small O radius) repulsive forces and the polarization forces that
tend to displace the ion from its position. Furthermore, the
on-site force constant of the O ion would become negative  and the
O ion would be unstable to the displacement from its regular high
symmetry site with local distortions. This would lead to the
instability of O similar to one observed for Li in LiNbO$_3$
~\cite{PhysRevB.61.272}. Indeed, although we found that the
so-called Bader volumes~\cite{bader:1990, Gonze:478} (which are a
measure of the ionic volumes) are smooth functions over the
pressure interval of interest, the Bader O volume, unlike those of
Pb and Ti, decreases faster under pressure than the volume of the
unit cell, as shown in Fig.~\ref{fig:Vol}. However, our \emph{ab
initio} lattice vibration calculations show that the on-site force
constant of O is {\it positive} at any pressures and, hence, the
local atomic potential possesses a single minimum. We thus can
rule out the drastic collapse of the oxygen ionic size as the
mechanism for the high-pressure ferroelectricity.

\subsection{Simple Tight-Binding Model}

To confirm the role of the interaction between the $2s$ orbitals
of oxygen atoms and the $d$ orbitals of the B atoms at high
pressure, we use the bond-orbital model of
Harrison~\cite{harrison:1980} -- a simplified tight-binding model,
where the Hamiltonian is limited to the on-site and
nearest-neighbor terms. This approximation allows an extraordinary
simplification of the problem and the total energy can be
rigorously calculated bond by bond. This is a very crude
approximation but, surprisingly, it works very well in a wide
variety of systems. In particular, it explains reasonably well the
dynamical charges in ferroelectric perovskites~\cite{ghosez:6224}.
In this section, we shall make use of the method to clarify the
role of the O $2s$ -- B $d$ and O $2p$ -- B $d$ couplings in the
high pressure ferroelectric instability in perovskites. We will
see that the instability at high pressure is governed by a
competition between short-range covalent and overlap interactions.

Intuitively, the gradually decreasing ferroelectricity with rising
hydrostatic pressure from zero to small values (i.e., before the
turning point) may be attributed to the fact that the conventional
mechanism responsible for triggering the ferroelectric instability
at zero pressure and related to the B $d$ -- O $2p$ hybridization
is not working well under pressure because of an increase of
covalency. It is known that in the case of significantly covalent
systems a paraelectric phase is of lower energy than the
ferroelectric one due to the increasing of the rigidity and the
stability of the metal-oxygen network~\cite{villesuzanne:307}.
From this point of view, it seems likely that -- besides the $pd$
covalency that appears to prevent the ferroelectric instability at
high pressure -- there is an interaction that reveals itself only
at high enough pressure (i.e., when the distance between atoms is
sufficiently small). This gives us a clue that the  interaction
should be small at large interatomic distances (e.g., at zero
pressure) and should increase more rapidly than covalency effects
at small interatomic distances, bringing the ferroelectric
instability into the system.

For a qualitative understanding of phase transformations under
pressure, we recall that two energy contributions to the cohesive
energy, namely the covalent and overlap energies, depend most
strongly on the interatomic distance. According to the Harrison
model, the (attractive) covalent interaction depends on the
interatomic distance, $d$, as $\sim d^{-7/2}$ while the
(repulsive) overlap interaction -- which is due to
nonorthogonality of wave functions on adjacent sites -- behaves as
$\sim d^{-7}$ and thus depends more strongly on the
nearest-neighbor distance. The interplay between the short-range
covalent and the overlap interactions can  be nicely seen from the
following arguments. The covalent energy contribution mainly
represents the short-range electron-ion potential energy while the
overlap energy can be predominantly ascribed to the increase of
the kinetic energy of the valence electrons upon
compression~\cite{PhysRevB.34.2787}. Since the resulting forces
due to these two interactions (central repulsive forces and
covalent forces) scale differently under pressure, one might
expect that at some pressure the noncentral nature of the bonding
will become negligible. Indeed, Fig.~\ref{fig:Cauchy} shows that
the high-pressure Cauchy relation between the elastic constants
for hydrostatic conditions $C_{12} = C_{44} + 2P$, where $C_{12}$
and $C_{44}$ are the elastic constants calculated from
first-principles, is satisfied for cubic PbTiO$_3$ in the vicinity
of the the turning point ($\simeq$ 30GPa), which indicates an
insignificant role of noncentral and many-body forces in this
pressure region. This fact implies that, for instance, empirical
two-body potentials could be used for calculations of elastic
properties for pressures around the turning point. Away from the
critical pressure the deviation from the Cauchy relation becomes
significant. The latter implies that the contribution of either
noncentral or many-body forces becomes more and more important at
higher pressures, and it cannot be treated anymore as a small
correction to the two-body potentials. Thus, we clearly see
manifestation of the competing forces under pressure.

Note that a major role of competition between covalent and overlap
interactions has been recognized for $sp$-bonded
solids~\cite{majewski:1366}. Remarkably, a drastic softening of
the transverse optical phonons across the pressure-induced phase
transition was predicted, and the physical origin of this
softening was shown to be closely related to
ferroelectricity~\cite{majewski:9666}.

Let us now proceed further by approximating, in Ti-based
perovskites, the total effect of the coupling (i) between the
oxygen $2s$ states and the $d$-orbitals of the Ti atom, and (ii)
between the oxygen $2p$ states and the $d$-orbitals of the Ti
atom, via the following expressions:

\begin{eqnarray}\label{eq:Eq1}
  E_{sd~bond} &=& 2V_3(sd)-2[V_2(sd)^2+ V_3(sd)^2]^{1/2}+E_{sd~rep}\\
  \nonumber
  E_{pd~bond} &=& 6V_3(pd)-6[V_2(pd)^2+ V_3(pd)^2]^{1/2}+E_{pd~rep}
\end{eqnarray}

where the (polar energies) $V_3$  are the average of the cation
and anion energy, and are given  by $V_3(sd)=(e_d-e_s)/2$ and
$V_3(pd)=(e_d-e_p)/2$ -- with $e_s=-29.14$ eV, $e_p=-14.13$ eV,
and $e_d=-11.04$ eV (which are the energies of the  $2s$ and  $2p$
states of the oxygen atom and of the $3d$ states of the titanium
atom, respectively~\cite{harrison:1980}. The second terms
$[...]^{1/2}$(which are the covalent energies) are connected with
the bond formation energy and are responsible for lowering the
energy of the bonding state. Based upon a second-moment
approximation~\cite{wills:4363}, these covalent energies are given
by $V_{2}(sd)^2 = \sum_j V_{sd\sigma}(r_j)^2$ and $V_{2}(pd)^2 =
\sum_j( V_{pd\sigma}(r_j)^2 +2V_{pd\pi}(r_j)^2 )/3$, where the
last sum is over the 6 nearest neighbors to Ti  and where the
so-called Harrison's two-center interactions, $V_{ldm}$, can be
taken to be of the form~\cite{harrison:1980}:

\begin{equation}\label{Vldm}
    V_{ldm} = \eta_{ldm}\frac{\hbar^{2}}{\mu}\frac{r_{d}^{3/2}}{r_j^{7/2}}
\end{equation}

where $r_j$ is the interatomic distance between $Ti$ and the
$j^{th}$ nearest-neighbor $O$, $\mu$ is the free-electron mass,
$r_d=1.029$ \AA~for titanium, $m=\sigma, \pi$, and $\eta$'s take
the following values: $\eta_{sd\sigma}=-3.16$,
$\eta_{sd\sigma}=-2.95$, and
$\eta_{sd\sigma}=1.36$~\cite{harrison:1980}.

The third terms of Eq.~(1), $E_{sd~rep}$ and $E_{pd~rep}$, are
related to the overlap interactions.  A repulsion comes from the
shift in the energy of each electron due to the nonorthogonality
of the $s$, $d$ and $p$, $d$ states on adjacent sites, with the
average shifts being given by ~\cite{PhysRevB.36.2695}:
\begin{eqnarray}
  \delta \varepsilon_{sd} &=& -\sum_{j} V_{sd\sigma}(r_j)S_{sd\sigma}(r_j) \\
  \nonumber
  \delta \varepsilon_{pd} &=& -1/3 \sum_{j}\left[
  V_{pd\sigma}(r_j)S_{pd\sigma}(r_j)+2V_{pd\pi}(r_j)S_{pd\pi}(r_j)
  \right ]
\end{eqnarray}
The overlap repulsion for a given bond,  is equal to this shift times the number of
electrons in these states. In the approximations adopted here we
use the extended H\"{u}ckel theory~\cite{hoffmann:1397} to derive
the overlap elements, as proposed by van Schilfgaarde and Harrison
~\cite{PhysRevB.33.2653}. It is plausible to take the overlap
elements to be proportional to the corresponding off-diagonal
hopping elements and inversely proportional to the on-site atomic
orbital energies: $S_{\alpha dm}(r) = \frac{2}{K_{\alpha
d}}V_{\alpha dm}(r)/(e_{\alpha} +e_{d})$, where $\alpha=s, p$ and
$m=\sigma, \pi$. For the Wolfsberg-Helmholz parameter $K_{\alpha
d}$, the
weighted formula $K_{\alpha d} = \kappa +
(e_\alpha-e_d)^2/(e_\alpha+e_d)^2 +
 (1 -\kappa) (e_\alpha-e_d)^4/(e_\alpha+e_d)^4$~\cite{ammeter:3686}
could be used. In general, the Wolfsberg-Helmholz parameter
$K_{\alpha d}$ can be given by the distance-dependent
form~\cite{calzaferri:11122}. For the sake of simplicity and
following Refs.~\cite{harrison:3592, majewski:9666,
baranowski:6287}, we assume that $K$ takes a different value for
each row of the Periodic Table, i.e., the $K$ of Carbon
($K_{C}=1.63$) for Oxygen and the $K$ of Ge ($K_{Ge}=1.07$) for
Titanium~\cite{harrison:1999} and take the average as
$K=(K_{C}K_{Ge})^{1/2}$.

Following this scheme, the covalent or polar bonding character of
a given bond (e.g., $sd$ or $pd$) can now be addressed. For each
bond, the bond polarity, $\alpha_p$, of the compound can be
obtained from $\alpha_p^2=V_3^2/(V_2^2+V_3^2)$ and a complementary
quantity, the covalency coefficient, $\alpha_c$, is defined
through the relation $\alpha_c^2=1-\alpha_p^2$. The tight-binding
model demonstrates the strong reduction of the {\it polarity} of
the $sd$ and $pd$ bonds under pressure, as shown in
Fig.~\ref{fig:Alphas}. On the other hand, the model predict a
considerable increase of the $sd~O-Ti$ chemical bond covalency
under pressure (i.e., when the interatomic distance decreases) --
which is consistent with the first-principles population analysis
reported in Section III.D.3. Indeed, as it can be clearly seen
from Fig.~\ref{fig:Alphas}, the $sd$ bond covalency depends more
strongly on the nearest-neighbor distance than the $pd$ bond
covalency, and therefore the first one may predominantly determine
the behavior of perovskites in the vicinity of the turning point.

In the Harrison bond-orbital formalism, the frequency of the $q=0$
optical phonon can be easily obtained~\cite{harrison:1980}.
Consider a perovskite structure viewed along a $\langle100\rangle
$ direction with a relative displacement $u_j=u_{Ti}-u_{O_j}$ of
the $Ti$ and $O$ atoms from their equilibrium positions in a
$\langle100\rangle $ direction when the atoms oscillate according
to a $\Gamma$-point soft mode. Accordingly, the change in length
of the indicated bond $Ti-O$ is $\delta d_j$, leading to a change
in the bonding energy of $\delta E_{sd~bond}$ and $\delta
E_{pd~bond}$. These changes in the bonding energy may be viewed as
the frozen-phonon energy obtained as the difference of the
respective energy relative to its equilibrium value, $\delta
E_{\alpha d~bond}=E_{\alpha d~bond}(u_j)-E_{\alpha d~bond}(0)$.
For the evaluation of the $sd$ and $pd$ contributions to the
frozen-phonon energy, usual harmonic fit function can be used,
i.e. $2~\delta E_{\alpha d~bond}=\omega^2_{\alpha d}\sum_{j} M_j
u_j^2$, where $M_j$ are the atomic masses, and $\omega_{\alpha d}$
are the contributions from $sd$ and $pd$ interactions to the total
frequency. [As an alternative to the above method, an analytic
expression for these changes can be obtained by expanding
Eq.~\ref{eq:Eq1} in a Taylor series with respect to the
displacements and then truncating the series to the second order
in the displacements.]

Let us consider the frequency of the $q=0$ optical mode in which
the $Ti$ and $O$ atoms move out of phase with the relative
displacements which reproduce a high-pressure eigenvector of the
soft mode obtained from first-principles (see, e.g., Fig.6). Since
above the turning point the relative displacement of each of the 4
transverse oxygens $u_{\bot}$ is greater than that of each of the
2 longitudinal oxygens $u_{\|}$, $2u_{\bot} > u_{\|}$, we have
chosen the following displacement pattern along the $z$ axis for
the 6 oxygen ions: $[0.4u,0.4u,u,u,u,u]$ with $u=0.002$. Denoting
$d$ as half of the cubic lattice constant We now see that one
neighbor distance becomes $d-0.4u$ (between the Ti atom and the
top O atom), one becomes $d+0.4u$ (between the Ti atom and the
bottom O atom) and four become $(d^2+u^2)^{1/2}$ (between the Ti
atom and 4 transverse O atoms).

The results of the frozen-phonon calculations are shown in
Fig.~\ref{fig:Froz}. At large interatomic distances both $sd$ and
$pd$ contributions stabilize the paraelectric state, since they
both have a {\it positive} $\delta E$. The curves of $~\delta
E_{sd~bond}$ and $~\delta E_{pd~bond}$ have maximums which lies at
around $2~\AA$ and $1.8~\AA$, respectively. One thus finds a
maximum at around $1.85~\AA$ for the sum of two contributions.
[Note that the experimental equilibrium interatomic distance in,
for instance, SrTiO$_3$ is $1.95~\AA$ and the turning point is
around $1.85~\AA$ according to our first-principles calculations.]
These estimates are compatible with the assumption that the
repulsion overlap energy does not become important until the
turning point. The most striking result is that the model predicts
a strong phonon softening upon compression since
Fig.~\ref{fig:Froz} shows that both $~\delta E_{sd~bond}$ and
$~\delta E_{pd~bond}$, that first increase with pressure and reach
a maximum at around the turning point, then decrease rapidly under
compression and become negative at $\approx1.83~\AA$ and
$\approx1.6~\AA$, respectively. The interatomic distance at which
the sum of both contributions becomes negative is
$\approx1.64~\AA$, to be compared with our \textit{ab-initio}
result of $\approx1.78~\AA$ for SrTiO$_3$. [It should be
emphasized that an exact agreement with \textit{ab-initio}
calculations cannot be expected, on account of the approximations
made in the tight-binding model, especially the neglect of
long-range interaction.] Interestingly, it is the $sd$
contribution which first destabilizes the paraelectric phase under
the influence of pressure. Therefore, the phonon frequency is
reduced under pressure after the turning point as much as this
destabilization is not compensated for by the other energy
contributions. Moreover and interestingly, as the system is
further compressed the $pd$ contribution also contributes to the
destabilizing of the parelectric phase. The present tigh-binding
model thus predicts a striking softening of a $\Gamma$-point
phonon mode under compression, and  clearly explains the strong
preference of perovskites to form ferroelectrically distorted
structures at high pressure due to an \emph{electronically driven}
structural instability (i.e., electron-phonon coupling). In
particular, this simple model reveals the predominant role played
by the costly-in-energy overlap between the $2s$ orbitals of
oxygen and the $3d$ orbitals of Ti on the high-pressure FE of
perovskites.

\subsection{Electronically Driven Instability}

According to the above simple tight-binding model, it is the
electrons and their interactions with ions (i.e., electron-phonon
coupling) that are responsible for the high-pressure ferroelectric
instability of perovskites. In this section, the electron-ion
interaction will be considered in more details. More precisely, we
shall be interested in the electronic contribution into the
ferroelectric soft mode within DFT and within the adiabatic
approximation.

Note that the adiabatic approximation consists in (a) neglecting
the phonon contributions to the electronic density response and
(b) approximating the dynamical response by a static
one~\cite{sham:301, geilikman:190}. In the adiabatic
approximation, the electrons are considered to remain in the
instantaneous (electronic) ground state when the lattice vibrates.
Consequently, the effect of electrons on the adiabatic phonon
frequencies is due to the fact that the change in the charge
density induced by the displacements of nuclei screens the
external perturbation (and the linear-response equations must be
solved self-consistently)~\cite{PhysRevB.54.16470}.

To refresh readers' memory, let us remind some basic equation of
density-functional theory (DFT). Within the framework of DFT, the
total energy can be expressed in terms of the Kohn-Sham
eigenvalues $\epsilon_i$:

\begin{equation}\label{eq:toten-eigv}
\quad E = \sum_i f_i\epsilon_i - {e^2\over 2}\int d\mathbf{r}
d\mathbf{r}' {n(\mathbf{r})n(\mathbf{r}')
\over|\mathbf{r}-\mathbf{r}'|} \hfill \hfill +
E_{xc}[n(\mathbf{r})] - \int n(\mathbf{r})
v_{xc}(\mathbf{r})d\mathbf{r} + E_N(\mathbf{R}).
\end{equation}
 where $f_i$ is the occupancy of the state $i$, and $n(\mathbf{r})$ is the
density of the valence electrons.  The first term is the band structure
energy:

\begin{equation}\label{eq:bs}
\sum_{i}f_i\epsilon_i=\sum_{i} f_i\langle i |
-1/2\mathbf{\nabla}^2 +V_{SCF}(\mathbf{r}) | i \rangle
\end{equation}
where an effective potential (also called {\it self-consistent},
SCF, potential) is:
\begin{equation} V_{SCF}(\mathbf{r}) = V_{ext}(\mathbf{r}) + e^2 \int {n(\mathbf{r}')\over|\mathbf{r}-\mathbf{r}'|}
d\mathbf{r}' + v_{xc}(\mathbf{r}), \label{eq:Vscf} \end{equation}
and $v_{xc}(\mathbf{r}) \equiv {\delta E_{xc}/ \delta
n(\mathbf{r})}$ is the exchange-correlation potential and
$V_{ext}(\mathbf{r})$ is the external (ionic) potential acting on
the electrons. $E_N(\mathbf{R})$ is the electrostatic interaction
between different nuclei:
\begin{equation}
\label{eq:ion-ion} E_N(\mathbf{R})= {e^2 \over 2} \sum_{I\ne J}
{Z_I Z_J \over |\mathbf{R}_I - \mathbf{R}_J | }.
\end{equation}

Note that, in general, the bare ion-ion contribution into phonon
modes can stabilize as well as destabilize the paraelectric state.
The destabilizing nature of this contribution is a trivial
consequence of Earnshaw's theorem~\cite{earnshaw:97, feynman:1975}
that states that a static system of electric charges can {\it not}
have a position of stable equilibrium. A a result, the crystals
stability is ensured by the electron contribution to the force
constants. Although it is possible that the ion-ion contribution
to a ferroelectric soft mode destabilizes this mode, it is not
necessarily the case -- implying that the ferroelectric soft mode
can be stable in a lattice of bare ionic
charges~\cite{kvyatkovskii:1}. This particular situation occurs in
the case of the ferroelectric soft mode in perovskites under
pressure, since we numerically found that the ionic term in the
force constants (which arises from the ion-ion Ewald
term~\cite{giannozzi:7231} with the bare ionic (pseudo)charges
(for example, $Z_{Pb}=14, Z_{Ti}=12, Z_{O}=6 $)) is positive in
the pressure range of interest.

Next, we are interested in the electronic contribution to the phonon
spectrum. The total second-order change in the electronic energy
$E^{(2)}_{el}$ can be written as~\cite{johnson:79,
PhysRev.188.1431}.

\begin{eqnarray}
  E^{(2)}_{el} &=& E^{(2)}_{el1}+E^{(2)}_{el2}\nonumber \\
  &=& \sum_{i\neq j} {f_i\langle i| \delta V_{ext}|j \rangle
  \langle j | \delta V_{ext}|i\rangle \over (\epsilon_i-\epsilon_j)} +
  \sum_{i} f_i\langle i| \delta V^{(2)}_{ext}|i\rangle \nonumber \\
   &=& 1/2\int d\mathbf{r}\delta n(\mathbf{r}) \delta
    V_{ext}(\mathbf{r})+
    \int d\mathbf{r}n^{0}(\mathbf{r})
    \delta V^{(2)}_{ext}(\mathbf{r})\label{eq:E2el}
\end{eqnarray}
where $n^{0}(\mathbf{r})$ is the electronic density of the
unperturbed crystal, $\delta n(\mathbf{r})$ is the change in the
electron density due to the ionic displacements, $\delta
V_{ext}(\mathbf{r})$ is the change in the  ``bare'' potential
experienced by the electron system to first order in the ionic
displacements, and $\delta V^{(2)}_{ext}(\mathbf{r})$ the second
order change.  Therefore, the expression for the electronic
contribution to the adiabatic force constant matrix consists of a
negative-definite contribution $E^{(2)}_{el1}$ (the first term in
Eq.~\ref{eq:E2el})  and a positive-definite contribution
$E^{(2)}_{el2}$ (the second term in
Eq.~\ref{eq:E2el})~\cite{kvyatkovskii:1}.

Let us now consider the negative-definite contribution
$E^{(2)}_{el1}$ to the adiabatic force constant matrix in more
details~\cite{geilikman:190, kvyatkovskii:1}. This contribution
describes how a first order change in the external Hamiltonian
leads to a first order change in the density matrix, which acts
back at the Hamiltonian.

Let us first consider the change in density $\delta n(\mathbf{r})$
induced by a change in the external potential $\delta
V_{ext}(\mathbf{r})$. Within DFT,
one can define the non-interacting electron polarizability,
$\chi_0(\mathbf{r},\mathbf{r}')$, as the charge-density response
to a variation of the total potential:
\begin{equation}
\delta n(\mathbf{r}) = \int d\mathbf{r}'\;
\chi_0(\mathbf{r},\mathbf{r}')\;\delta V_{SCF}(\mathbf{r}') .
\label{eq:chi0def}
\end{equation}
The expression for $\chi_0(\mathbf{r},\mathbf{r}')$ has the well
known form~\cite{pick:910, PhysRev.188.1431, sham:301}:
\begin{equation}
\chi_0(\mathbf{r},\mathbf{r}') = \sum_{i,j}
\frac{f_i-f_j}{\epsilon_i-\epsilon_j}
              \;\;  \psi^*_i(\mathbf{r}) \psi_j(\mathbf{r})
              \psi^*_j(\mathbf{r}') \psi_i(\mathbf{r}')
\label{eq:chi0}
\end{equation}
where the sums over $i$ and $j$ extend to both occupied and empty
states. Clearly, only terms involving virtual transitions from
occupied or partially occupied to empty or partially empty states
contribute because of the difference $f_i-f_j$ in the numerator.

Recall the relationship between the bare and the self-consistent
perturbing potential:
\begin{equation}
     \delta V_{SCF}(\mathbf{r}) =
        \delta V_{ext}(\mathbf{r}) + \int d\mathbf{r}'\;
        K(\mathbf{r},\mathbf{r}')\delta n(\mathbf{r}') ,
\label{eq:DeltaV}
\end{equation}

where: \begin{equation} K(\mathbf{r},\mathbf{r}') = {e^2\over
|\mathbf{r}-\mathbf{r}'|} + {\delta^2 E_{xc} \over \delta
n(\mathbf{r}) \delta n(\mathbf{r}') }, \label{eq:Kappa}
\end{equation}

Combining Eqs.~(\ref{eq:E2el}--\ref{eq:DeltaV}), one can rewrite
the the negative-definite contribution as follows:

\begin{eqnarray}
  E^{(2)}_{el1}&=&
  1/2 \int d\mathbf{r}~\delta n(\mathbf{r}) \delta V_{ext}(\mathbf{r})\nonumber \\
  &=& 1/2 \int d\mathbf{r}d\mathbf{r}'\delta V_{SCF}(\mathbf{r})
    \chi_0(\mathbf{r},\mathbf{r}') \delta
    V_{SCF}(\mathbf{r}')  - 1/2 \int d\mathbf{r}'~\delta n(\mathbf{r})
    K(\mathbf{r},\mathbf{r}')\delta n(\mathbf{r}')
    \label{eq:E2elN}
\end{eqnarray}

Substituting the results (\ref{eq:Kappa}) and (\ref{eq:E2elN}) in
Eq.~(\ref{eq:E2el}) we obtain

\begin{eqnarray}
  E^{(2)}_{el}
  &=& 1/2\int d\mathbf{r}d\mathbf{r}'\delta V_{SCF}(\mathbf{r})
    \chi_0(\mathbf{r},\mathbf{r}') \delta
    V_{SCF}(\mathbf{r}')+
    \int d\mathbf{r}n^{0}(\mathbf{r})
    \delta V^{(2)}_{ext}(\mathbf{r})\nonumber \\
    &-&1/2\int d\mathbf{r}'~\delta n(\mathbf{r})
    \left[ {e^2\over |\mathbf{r}-\mathbf{r}'|} + {\delta^2 E_{xc} \over \delta
    n(\mathbf{r}) \delta n(\mathbf{r}') } \right ] \delta n(\mathbf{r}')
    \label{eq:E2el2}
\end{eqnarray}

The first two terms in Eq.~(\ref{eq:E2el2}) are due to the
variation of the band energy term in the total energy (the first
term in Eq.~(\ref{eq:toten-eigv})) and the third term in
Eq.~(\ref{eq:E2el2}) is due to the variation of the Hartree energy
and the exchange-correlation energy.

The total electronic contribution (see Eq.~\ref{eq:E2el}) to the
interatomic force constant tensor $\Phi^{\alpha\beta}_{ll'}$
between the atoms at $\mathbf{r}_l$ and $\mathbf{r}_{l'}$ can be
linked to the second-order change in the electronic energy with
respect to collective ionic displacements in the usual way as
$2E^{(2)}_{el} = \sum_{l,l'}
\Phi^{\alpha\beta}_{ll'}u_{l,\alpha}u_{l',\beta}$, where
$u_{l,\alpha}$ is the displacement of the $l$th ion in the
$\alpha$-direction from its equilibrium position $\mathbf{R}_l$.

Taking into account that $\delta V_{SCF}(\mathbf{r}) = \sum_{l}
{\delta V_{SCF}(\mathbf{r}) / \delta \mathbf{R}_l} \cdot
\mathbf{u}_l$,  the negative-definite contribution of the band
structure energy $E^{(2)}_{el1}$ to the dynamical matrix  can be
written in the form~\cite{PhysRevB.19.6142}

\begin{equation}\label{Dyn}
    \Phi^{\alpha\beta}_{ll'}(\mathbf{q}) = {1\over N_c}\sum_{\lambda\lambda',\mathbf{k}}
\frac{f_{\mathbf{k}+\mathbf{q}\lambda'}-f_{\mathbf{k}\lambda}}
{\epsilon_{\mathbf{k}+\mathbf{q}\lambda'}-\epsilon_{\mathbf{k}\lambda}}
              \;\;  g^{l\alpha
              \ast}_{\mathbf{k}+\mathbf{q}\lambda',\mathbf{k}\lambda}
              g^{l'\beta}_{\mathbf{k}+\mathbf{q}\lambda',\mathbf{k}\lambda}
\end{equation}\

where $N_c$ the number of unit cells in the crystal, and
 $g^{l}_{\mathbf{k}'\lambda',\mathbf{k}\lambda}=
\langle\mathbf{k}'\lambda'|\delta V_{SCF}(\mathbf{r}) / \delta
\mathbf{R}_{l\mathbf{q}}|\mathbf{k}\lambda\rangle$ are the
electron-phonon matrix-elements. The sum has to be carried out,
 in principle,  over all bands
$\lambda$,$\lambda'$ and all $k$ points $\mathbf{k},~\mathbf{k}'$
that can be connected via the phonon wave vector $\mathbf{q}$.
Interestingly, simplifying the electron-phonon matrix-elements
within nonorthogonal tight-binding representation
as~\cite{PhysRevB.19.6142, PhysRevLett.39.1094}
$g_{\mathbf{k}\lambda,\mathbf{k}'\lambda'} \propto {(
\partial \epsilon_{\mathbf{k}\lambda}/
\partial\mathbf{k}
[A^{\dag}_{\mathbf{k}}S_{\mathbf{k}}A^{\dag}_{\mathbf{k'}}]_{\lambda\lambda'}
- \partial \epsilon_{\mathbf{k}'\lambda'}/
\partial\mathbf{k}'
[A^{\dag}_{\mathbf{k}}S_{\mathbf{k'}}A^{\dag}_{\mathbf{k'}}]_{\lambda\lambda'})}$
-- where $S$ is the overlap  and  $A$ the eigenvector matrix -- we
clearly see the dependence of the electron-phonon interaction on
the overlap effects. An important feature of this expression noted
in Refs.~\cite{PhysRevB.19.6142, PhysRevLett.39.1094} is the fact
that the electron-phonon coupling is large where the electron
velocities are large and therefore the density of states are
small. Indeed, we numerically checked these facts in our first
principles calculations.

Interestingly, Eq.~(\ref{Dyn}) is in fact the electron-phonon
coupling in the so-called Fr\"{o}hlich model~\cite{frohlich:325}
-- which thus proves that such coupling is naturally included in
the adiabatic approximation within DFT~\cite{geilikman:190,
kvyatkovskii:1}. (Note that the electron-phonon coupling is
responsible for the ferroelectric instability in the vibronic
theory of ferroelectricity~\cite{3274575}, i.e., the existence of
the pseudo Jahn-Teller effect due to the existence of two closely
situated filled $p$ and $d$ empty bands in perovskites-like
structures~\cite{1964A17519}).

Eq.~(\ref{Dyn}) has been used to calculate the electron-phonon
contribution to the ferroelectric soft mode at the $\Gamma$ point
in PbTiO$_3$ and NaNbO$_3$. More precisely, we obtained the
eigenvalues of the LDA-determined
$\Phi^{\alpha\beta}_{ll'}(\mathbf{q})$ matrix corresponding to the
LDA-determined eigenvectors of the total dynamical matrix. The
resulting eigenvalue $\omega^2_{ep}$ corresponding to the soft TO
mode is negative for any pressure and shown in Fig.~\ref{fig:D2},
and clearly behaves in a nonmonotonic way with pressure.
Interestingly, this nonmonotonic behavior varies in a correlated
way with the soft-mode total frequency (see
Figs.~\ref{fig:Omega3}, \ref{fig:OmegaAll}, \ref{fig:OmegaWO})
under pressure-- which thus confirms that the electron-phonon
coupling is the leading mechanism for the high-pressure
ferroelectricity.

\section{Concluding Remarks.}

In summary, our results reveal that, unlike commonly thought,
ferroelectricity is not suppressed by high pressure in insulating
perovskites. Instead, ferroelectricity is found to enhance as
pressure increases above a critical value. Moreover, this
unexpected high-pressure ferroelectricity is different in nature
from conventional ferroelectricity because it is driven by an
original electronic effect rather by long-range ionic
interactions. Depending on the energetic difference between
ferroelectric and paraelectric phases at atmospheric pressure,
such phenomenon can lead to various overlooked/original effects,
e.g. (i) the disappearance and then re-entrance of
ferroelectricity under pressure; (ii) the occurrence of FE at high
pressure in a nominally-paraelectric compound; and (iii) the
existence of FE at any pressure (see Figure 6). We thus hope that
this article, and its numerous details, will stimulate the
investigation of novel effects in ``smart'' materials under
pressure.

We would like to thank S. Prosandeev and I. Ponomareva for useful
discussions. This work is supported by ONR grants
N00014-01-1-0365, N00014-04-1-0413 and N00014-01-1-0600, by NSF
grants DMR-0404335, and by DOE grant DE-FG02-05ER46188.

\newpage

\begin{figure}
\includegraphics[width=0.7\textwidth]{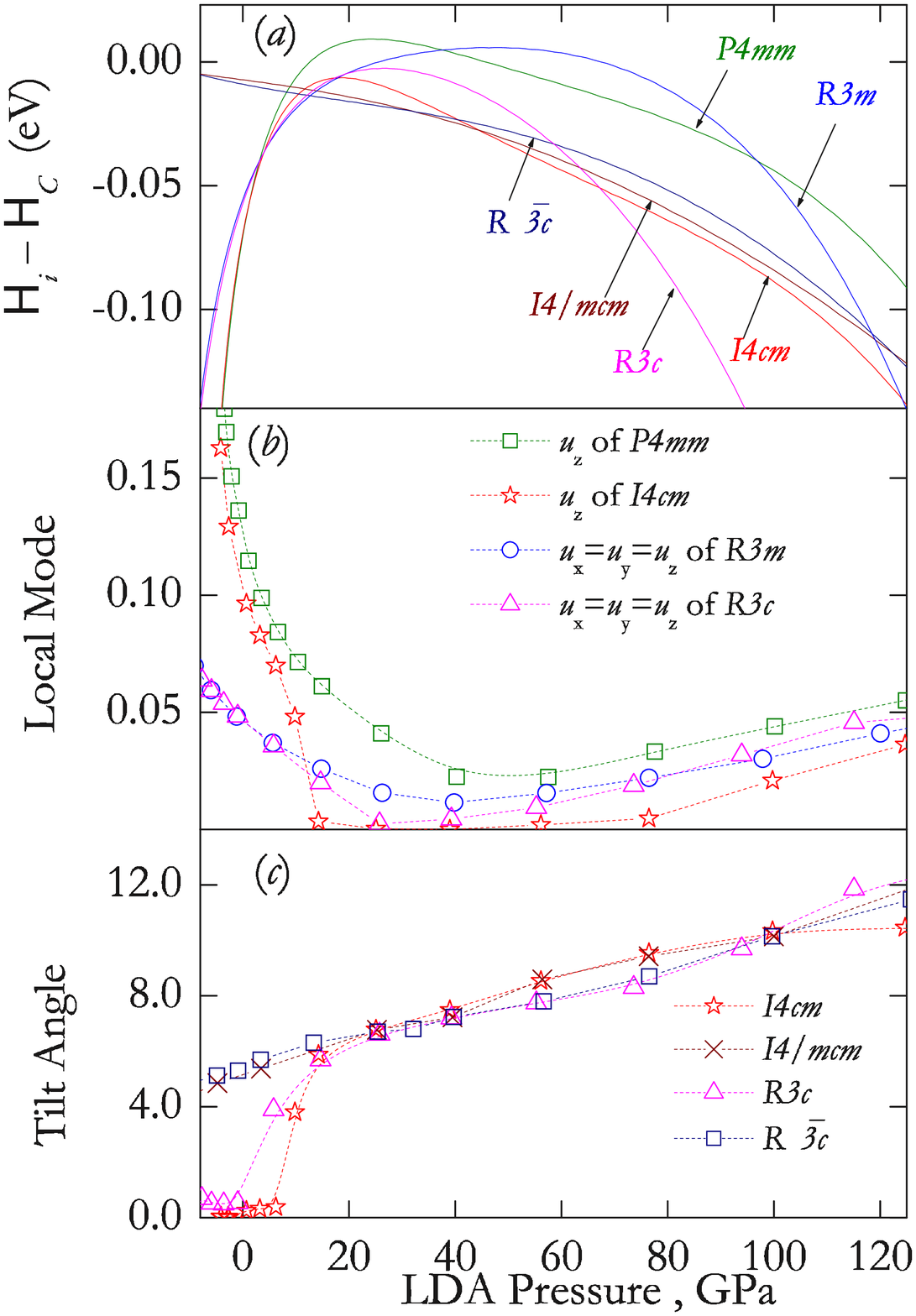}
\caption{First-principles prediction of the pressure behavior of
(a) $\Delta \mathcal{H}$ (see text) for all the considered phases
(the phase corresponding to a minimal  $\Delta \mathcal{H}$ at a
given $P$ is thus the most stable one for this pressure), (b) the
Cartesian components (along the pseudo-cubic $\langle 001 \rangle$
axes) of the polar local soft-mode~\cite{zhong:6301} - which is
directly proportional to the spontaneous polarization - for the
P4mm, I4cm, R3m and R3c phases, and (c) the rotational angle of
the oxygen octahedra with respect to the pseudo-cubic $\langle 111
\rangle$ direction for the R3c and R$\overline{3}$c rhombohedral
phases and with respect to the pseudo-cubic $\langle 001 \rangle$
direction for the I4cm and I4/mcm tetragonal
phases.}\label{fig:Fig1}
\end{figure}

\begin{figure}
\includegraphics[width=1\textwidth]{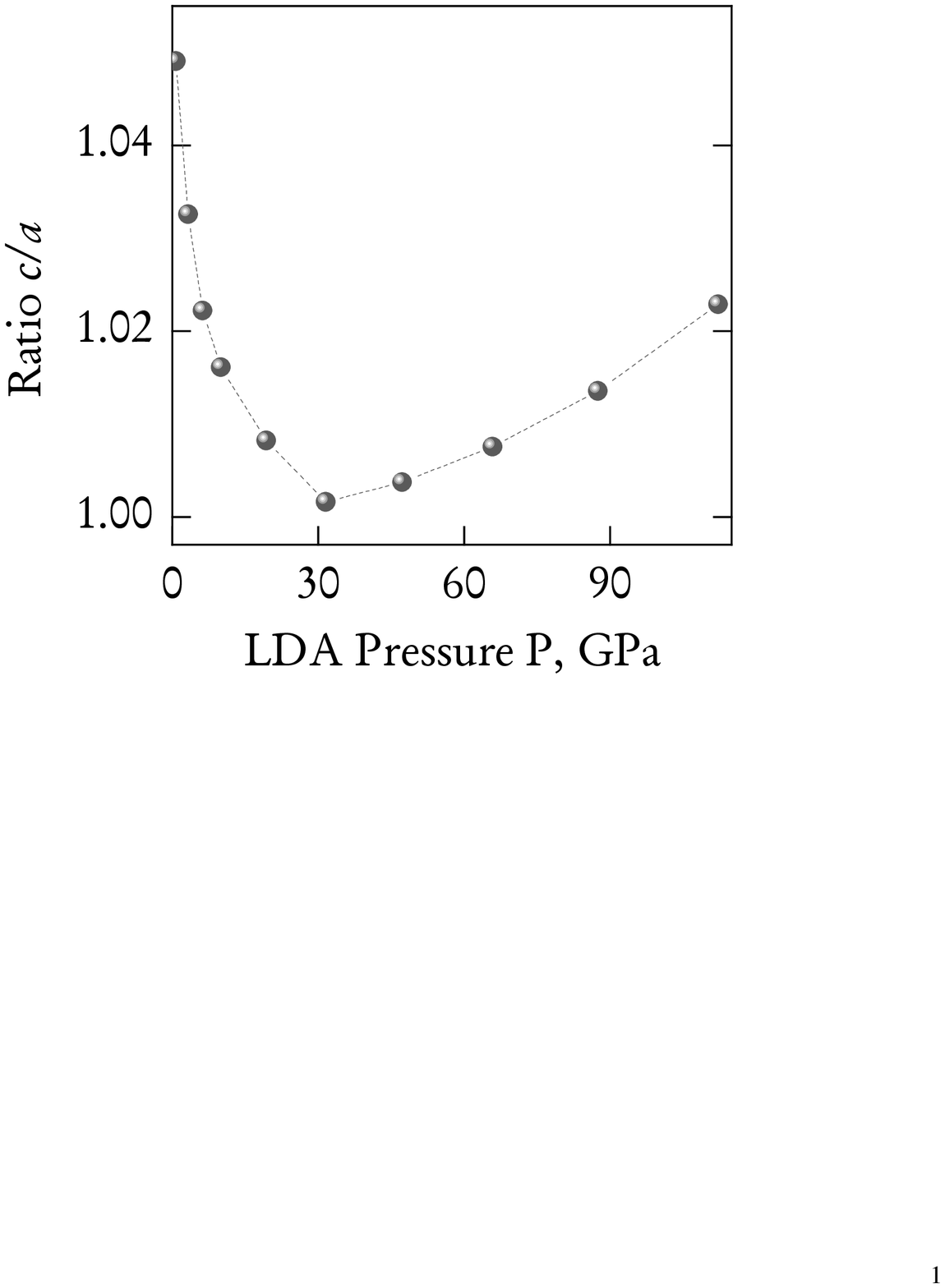}
\caption{ The LDA-predicted pressure behavior of the $c/a$
distortion at 0K for the P4mm phase in
PbTiO$_3$.}\label{fig:FigCA}
\end{figure}

\begin{figure}
\includegraphics[width=1\textwidth,height=0.75\textheight]{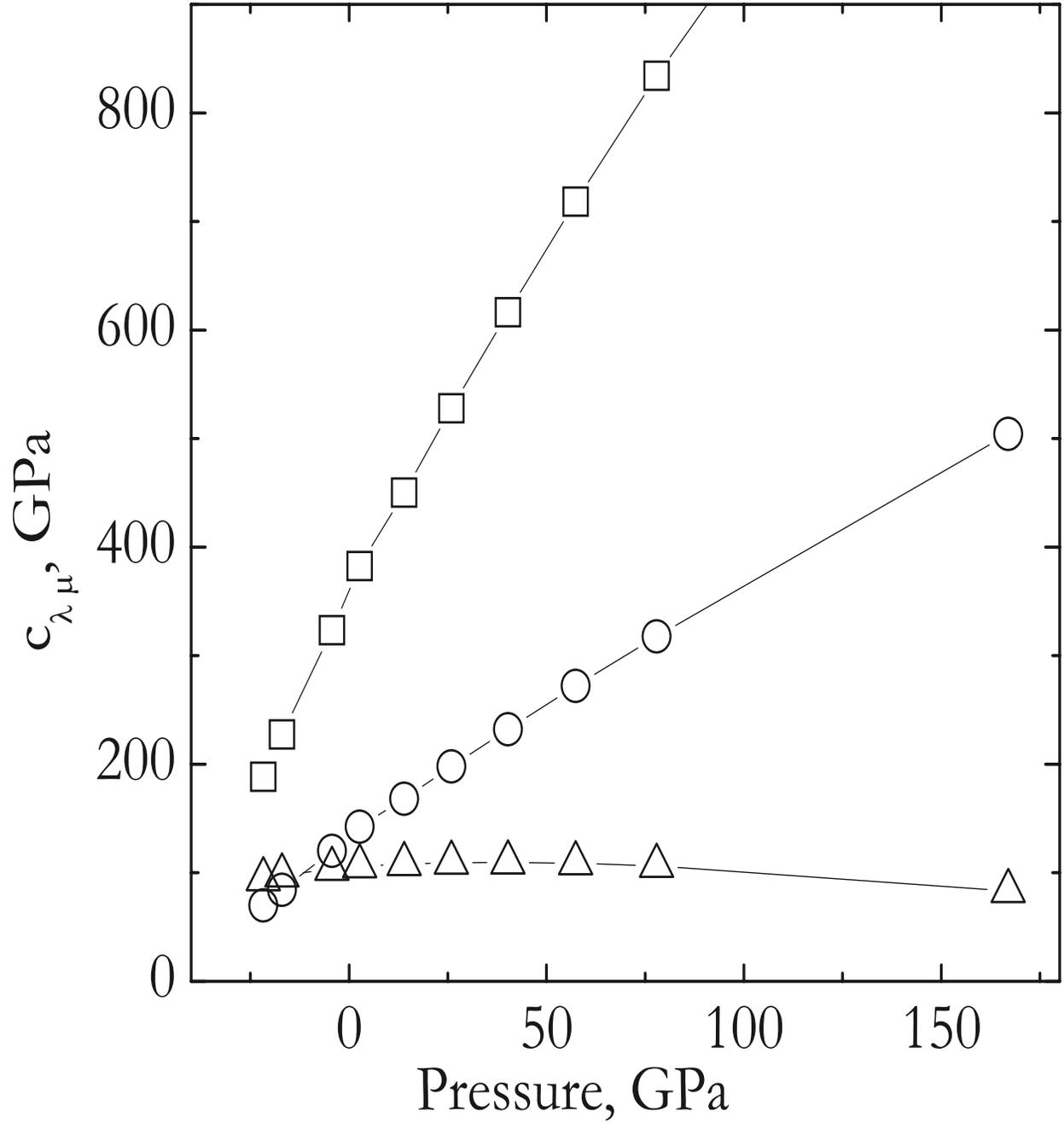}
\caption{The pressure dependence of the elastic stiffness
coefficients $c_{11}$ (squares), $c_{12}$ (dots) , and $c_{44}$
(triangles) of the cubic phase of PbTiO$_3$, as predicted by LDA.
}\label{fig:Elast}
\end{figure}

\begin{figure}
\includegraphics[width=1\textwidth,height=.75\textheight]{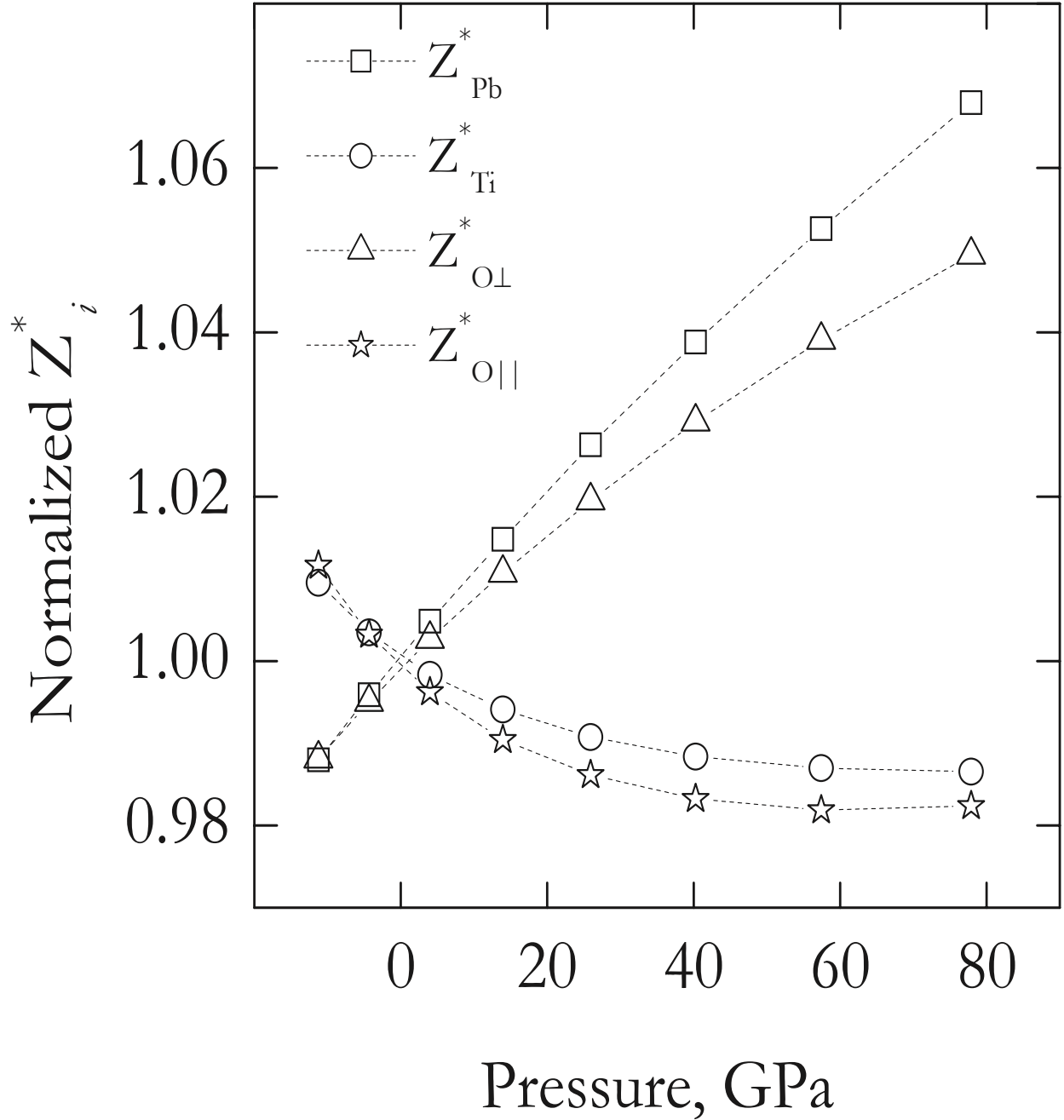}
\caption{Pressure evolution of the normalized Born effective
charges for Pb, Ti, O$_{\perp}$ and O$_\parallel$ in the cubic
phase of PbTiO$_3$. The normalization is done with respect to the
Born effective charges at zero pressure obtained with ABINIT for
Pb, Ti, O$_\perp$, and O$_\parallel$ in the cubic
$Pm\overline{3}m$ phase -- that are 3.915, 7.14, -2.61 and  -5.85,
respectively.}\label{fig:ZB}
\end{figure}

\begin{figure}
\includegraphics[width=1\textwidth,height=.95\textheight]{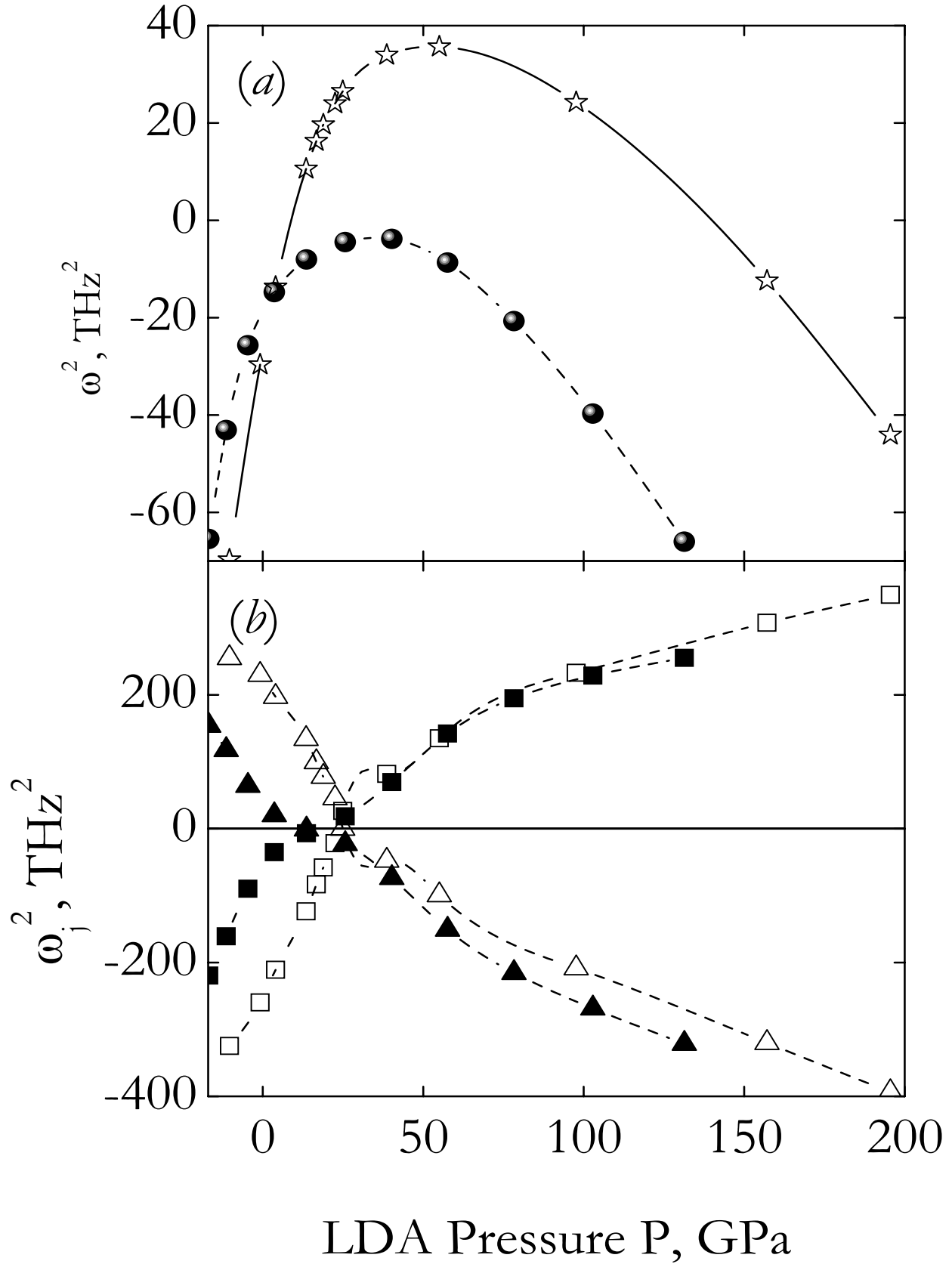}
\caption{Pressure behavior of phonons in PbTiO$_{3}$ (solid
symbols) and BaTiO$_{3}$ (open symbols). Panel (a) displays the
square of the zone-center TO phonon frequency in the cubic phase
within LDA at 0K. Panel (b) shows the contributions from the
Coulomb interactions (squares) within the rigid ions model and
non-Coulomb interactions (triangles) to this
square.}\label{fig:Omega}
\end{figure}

\begin{figure}
\includegraphics[width=1\textwidth,height=.55\textheight]{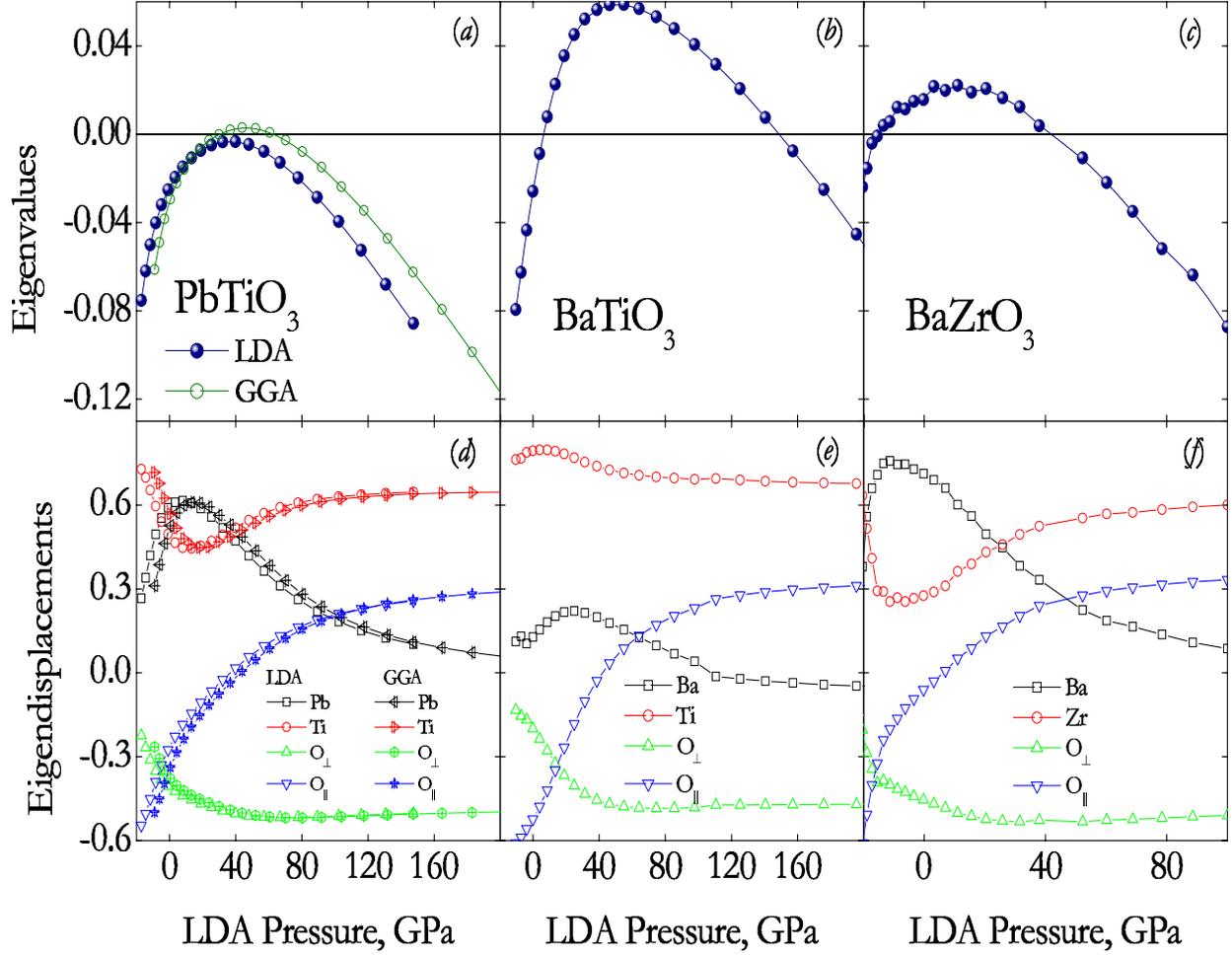}
\caption{ LDA-predicted pressure behavior of the
soft-mode eigenvalue and eigenvectors of the second-derivative
matrix in cubic PbTiO$_3$ (panels (a) and (d)), BaTiO$_3$ (panels
(b) and (e)) and BaZrO$_3$ (panels (c) and (f)). The GGA results
are also shown for PbTiO$_3$ in panels (a) and (d), and suggest
that PbTiO$_3$ may be paraelectric within a small pressure range.
O$_{3}$ is the oxygen atom located between two (slightly
displaced) B-atoms along the [001] direction.}\label{fig:Omega3}
\end{figure}

\begin{figure}
\includegraphics[width=1\textwidth,height=0.35\textheight]{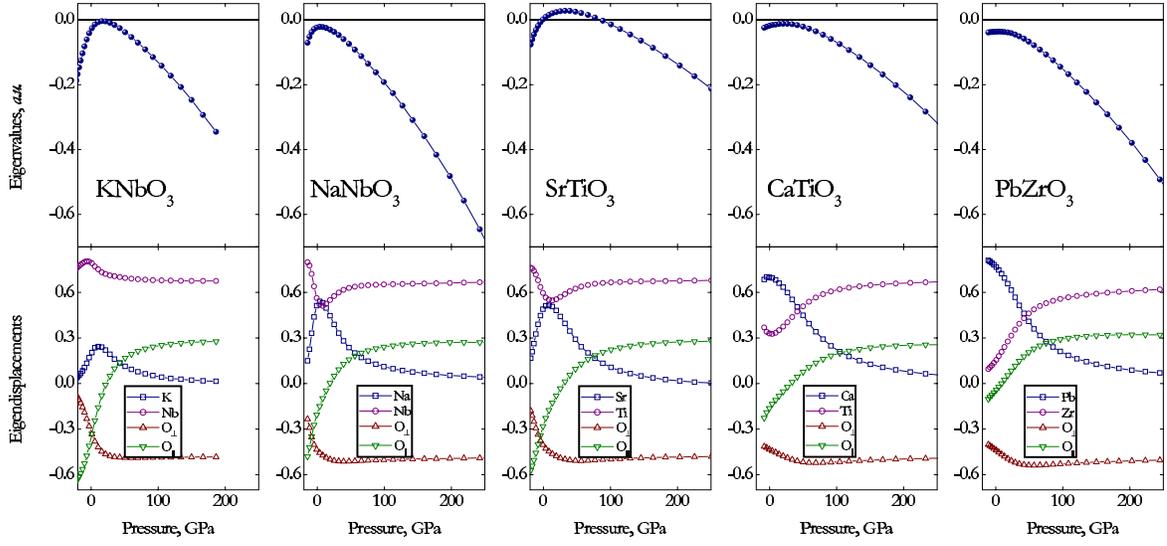}
\caption{ LDA-predicted pressure behavior of the soft-mode
eigenvalue and eigenvectors of the second-derivative matrix in
cubic KNbO$_3$, NaNbO$_3$, SrTiO$_3$, CaTiO$_3$ and
PbZrO$_3$.}\label{fig:OmegaAll}
\end{figure}

\begin{figure}
\includegraphics[width=1\textwidth,height=.75\textheight]{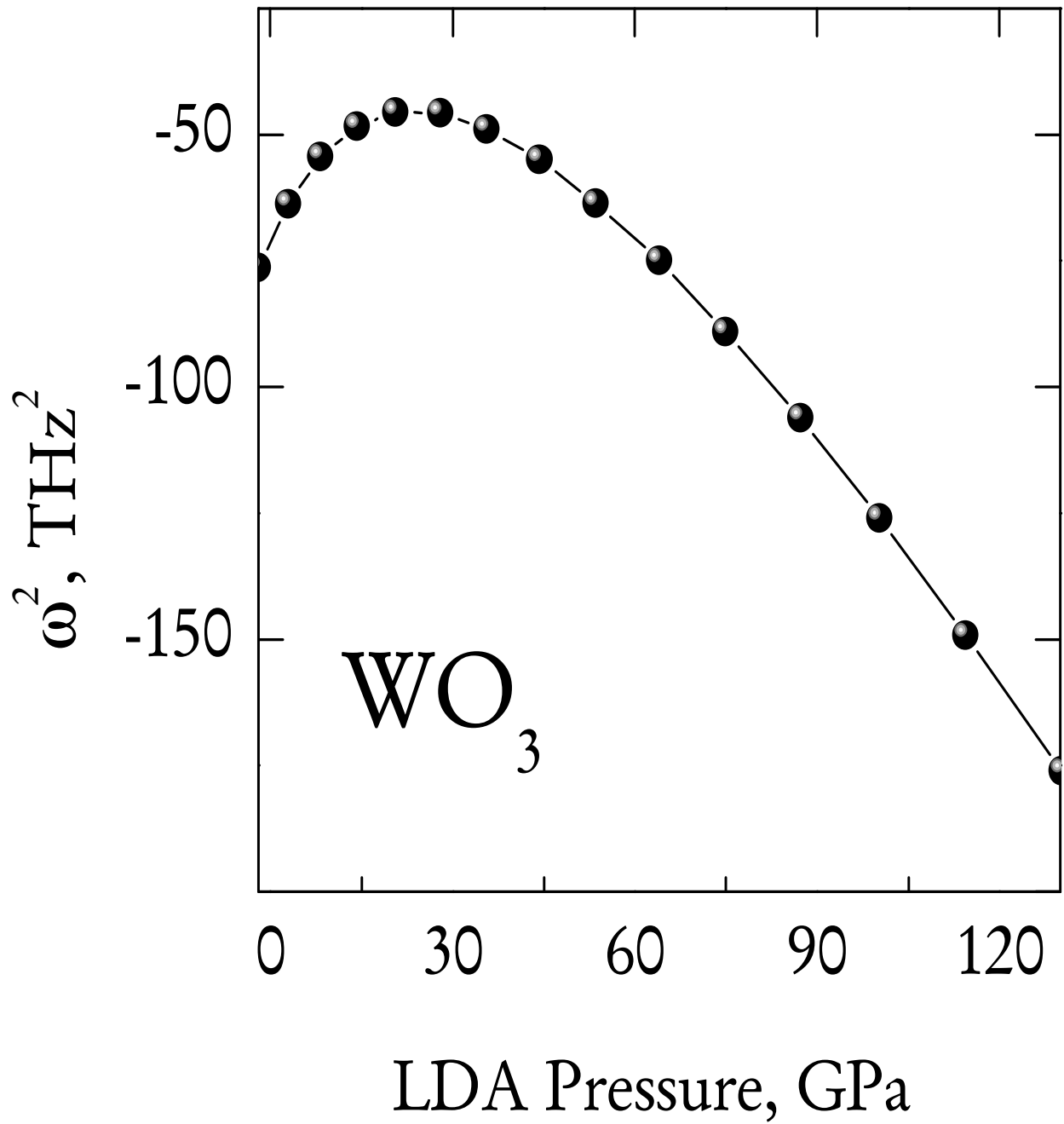}
\caption{ LDA-predicted pressure behavior of the soft-mode frequency square in
cubic WO$_3$.}\label{fig:OmegaWO}
\end{figure}

\begin{figure}
\includegraphics[width=1\textwidth,height=.75\textheight]{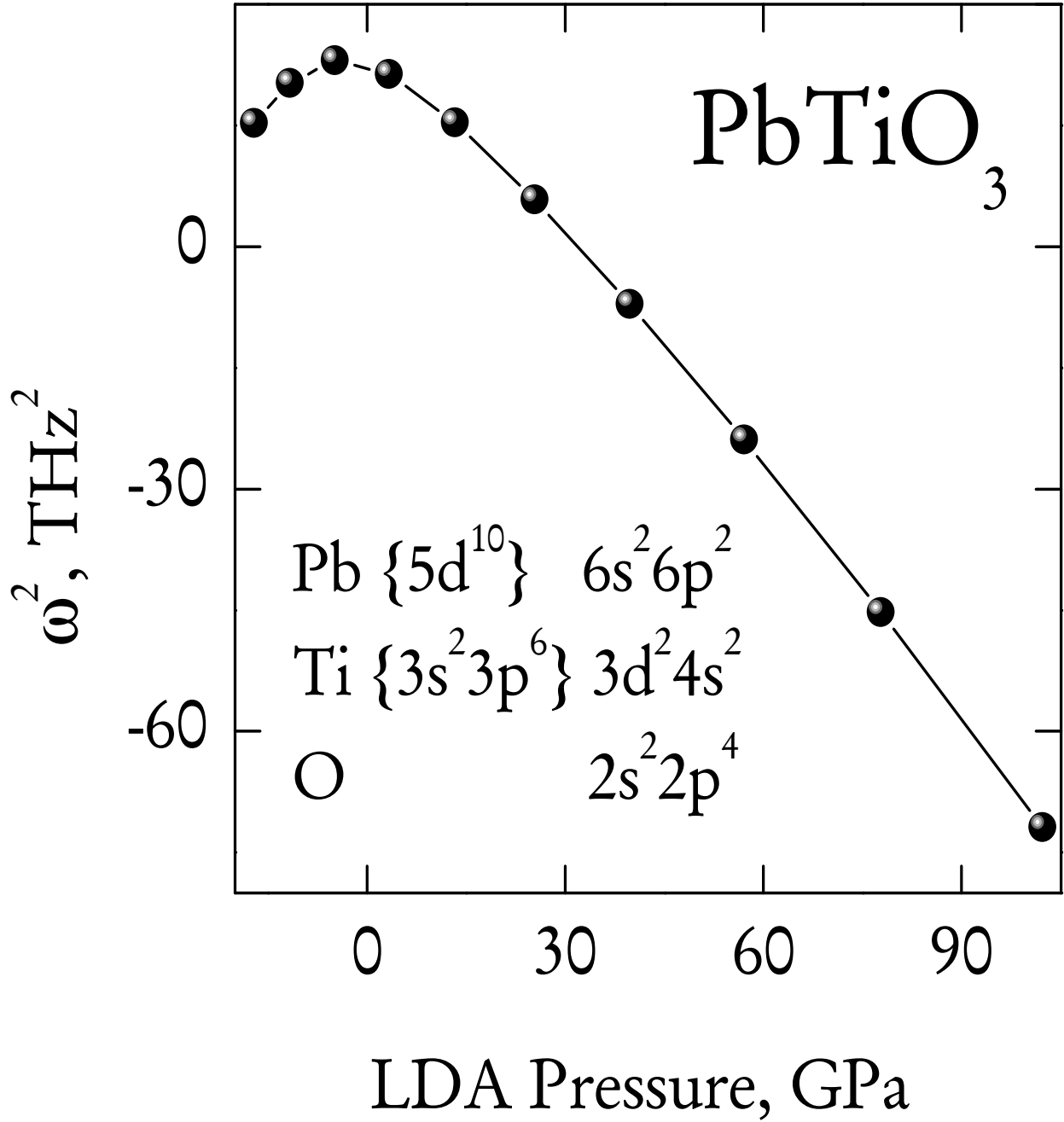}
\caption{LDA-predicted pressure behavior of the soft-mode
frequency square in the cubic phase of PbTiO$_3$, when excluding
the 5d semicore states of Pb and 3s and 3p semicore states of Ti
from the valence electrons (unlike the 6s and 6p states of Pb, the
3d and 4s states of Ti and the 2s and 2p states of oxygen that are
kept in the valence).} \label{fig:FigSCr}
\end{figure}

\begin{figure}
\includegraphics[width=1\textwidth,height=.45\textheight]{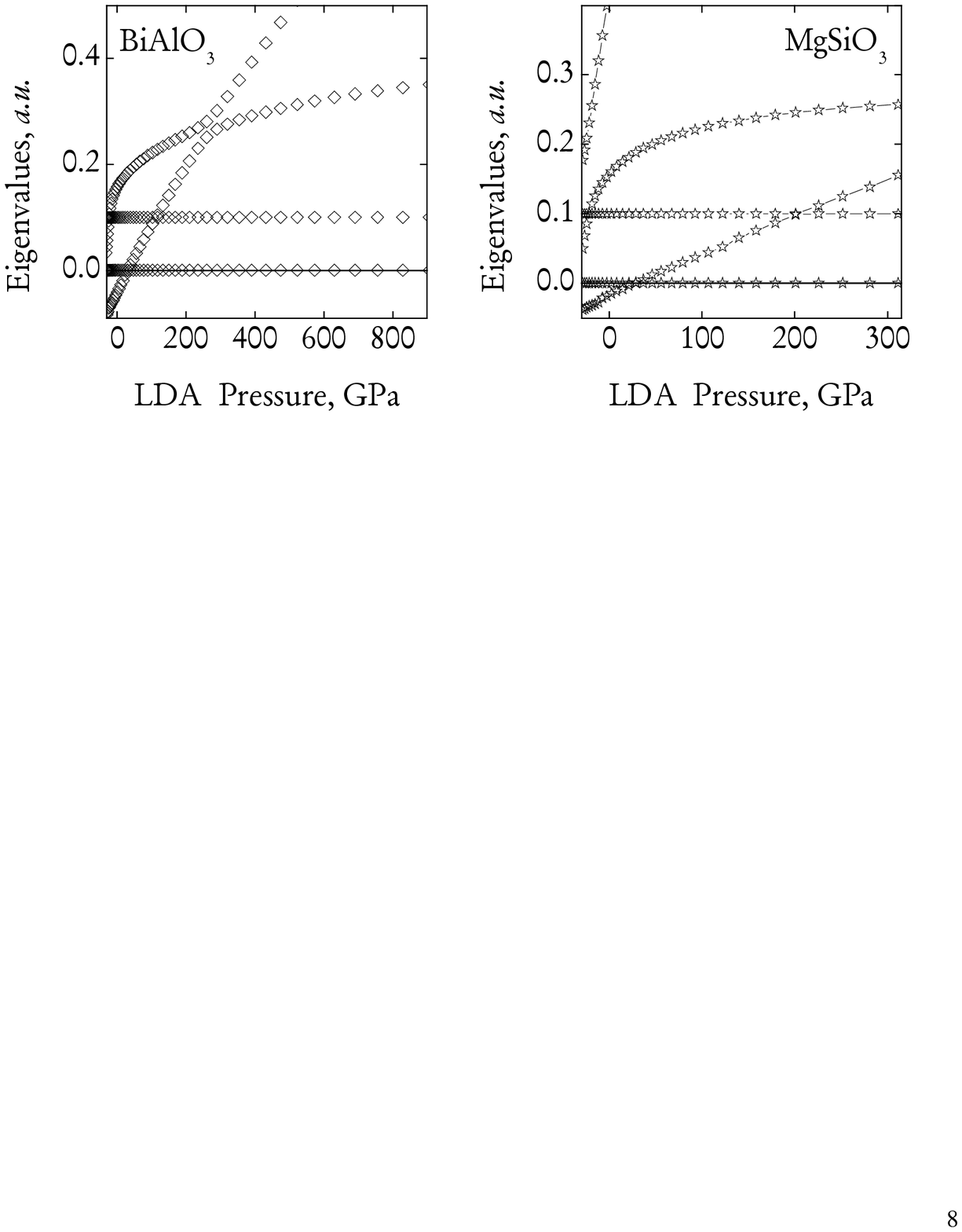}
\caption{ LDA-predicted pressure behavior of the soft-mode
eigenvalues of the second-derivative matrix in cubic bismuth
aluminate BiAlO$_3$ and magnesium silicate MgSiO$_3$.}
\label{fig:FigNOd}
\end{figure}

\begin{figure}
\includegraphics{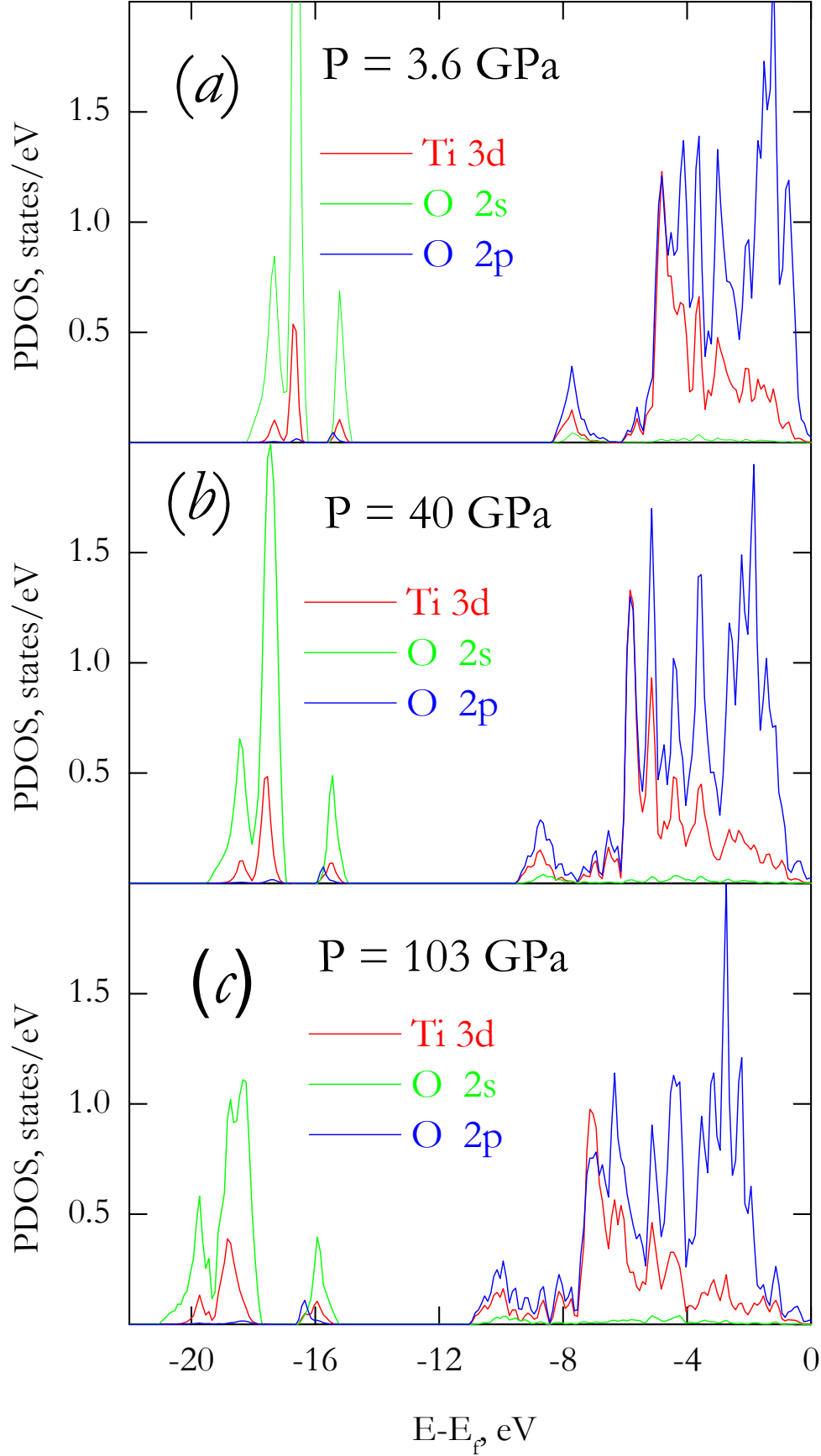}
\caption{Electronic-related properties of PbTiO$_{3}$ under
pressure. Panels (a), (b) and (c) display the partial electronic
density of occupied states in the cubic phase at 3.6, 40 and 103
GPa, respectively, for the O \textit{{\small 2s}}, O
\textit{{\small 2p}} {\small and Ti} \textit{{\small 3d}}
orbitals. The zero in energy is chosen at the top of the valence
band, E$_f$.}\label{fig:PTOdos} \end{figure}

\begin{figure}
\includegraphics[width=1\textwidth,height=.65\textheight]{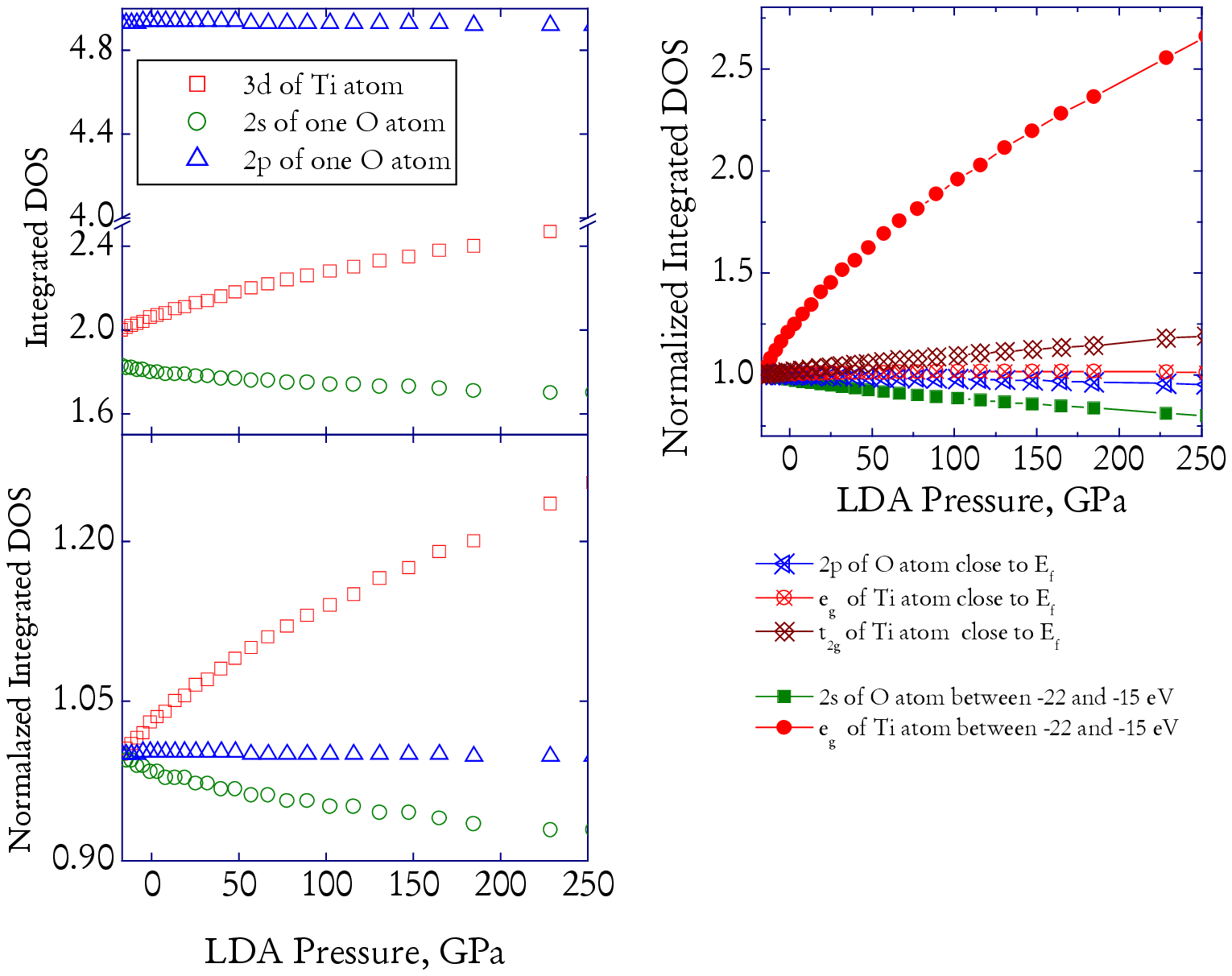}
\caption{Integrated DOS of PbTiO$_{3}$ under pressure in the cubic
state. Panel (a) displays the integrated PDOS for Ti \textit{3d}
and O \textit{2p} and \textit{2s} states; panel (b) displays the
normalized integrated PDOS for the same states; panel (c) displays
the contributions to the normalized integrated PDOS coming from
the low-lying O \textit{2s} and Ti \textit{3d} $e_g$ states
(located between -22 and -15 eV) and from the topmost group of the
valence O \textit{2p} and Ti \textit{3d}
bands.}\label{fig:ipdosPTO1}
\end{figure}

\begin{figure}
\includegraphics[width=1\textwidth,height=.75\textheight]{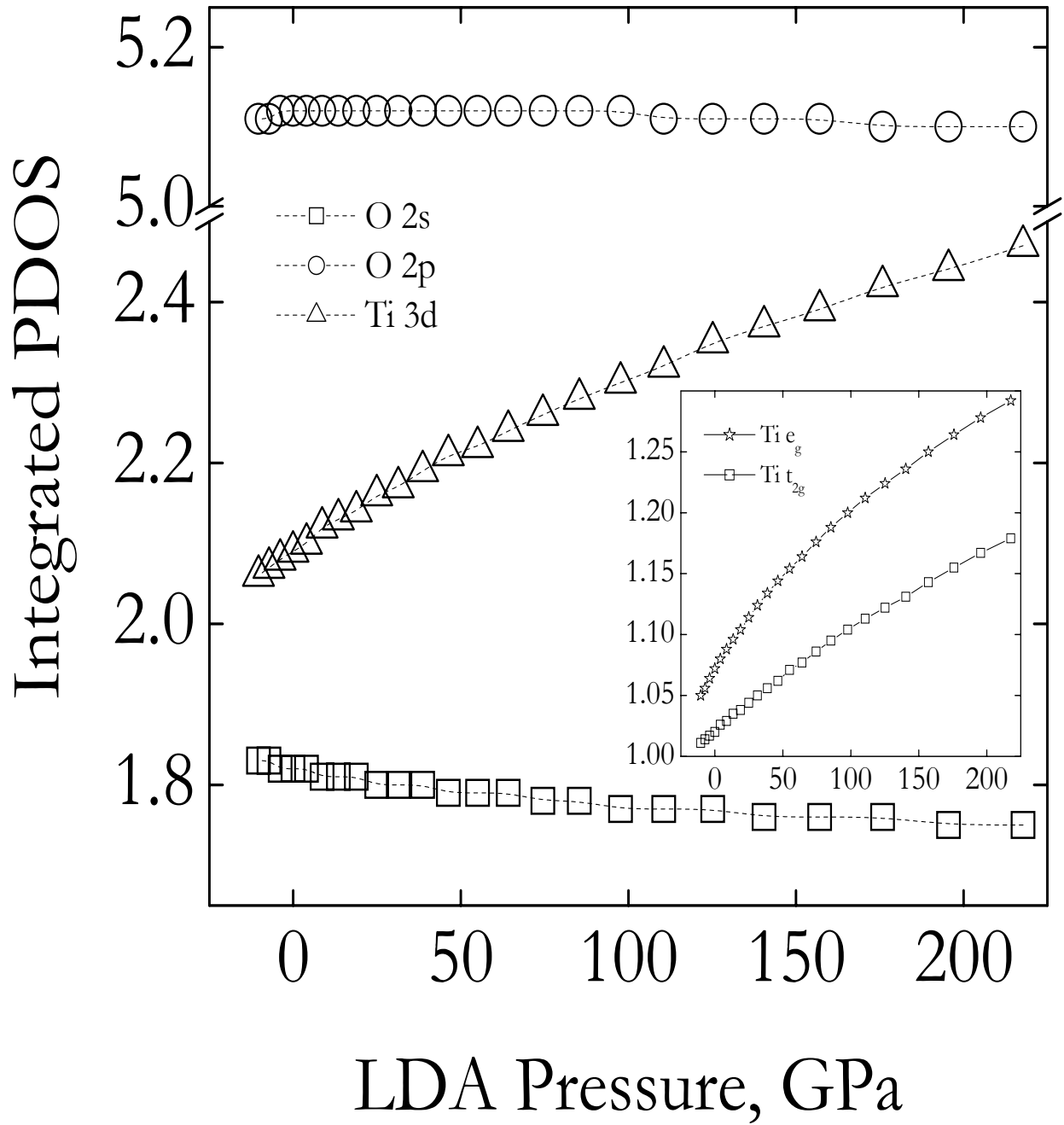}
\caption{Integrated PDOS of BaTiO$_{3}$ under pressure in the
cubic state. The inset shows the variation of the integrated PDOS
of Ti \textit{3d} $e_g$ and Ti \textit{3d} $t_{2g}$ versus
pressure. }\label{fig:ipdosBTO1}
\end{figure}

\begin{figure}
\includegraphics[width=1\textwidth,height=.75\textheight]{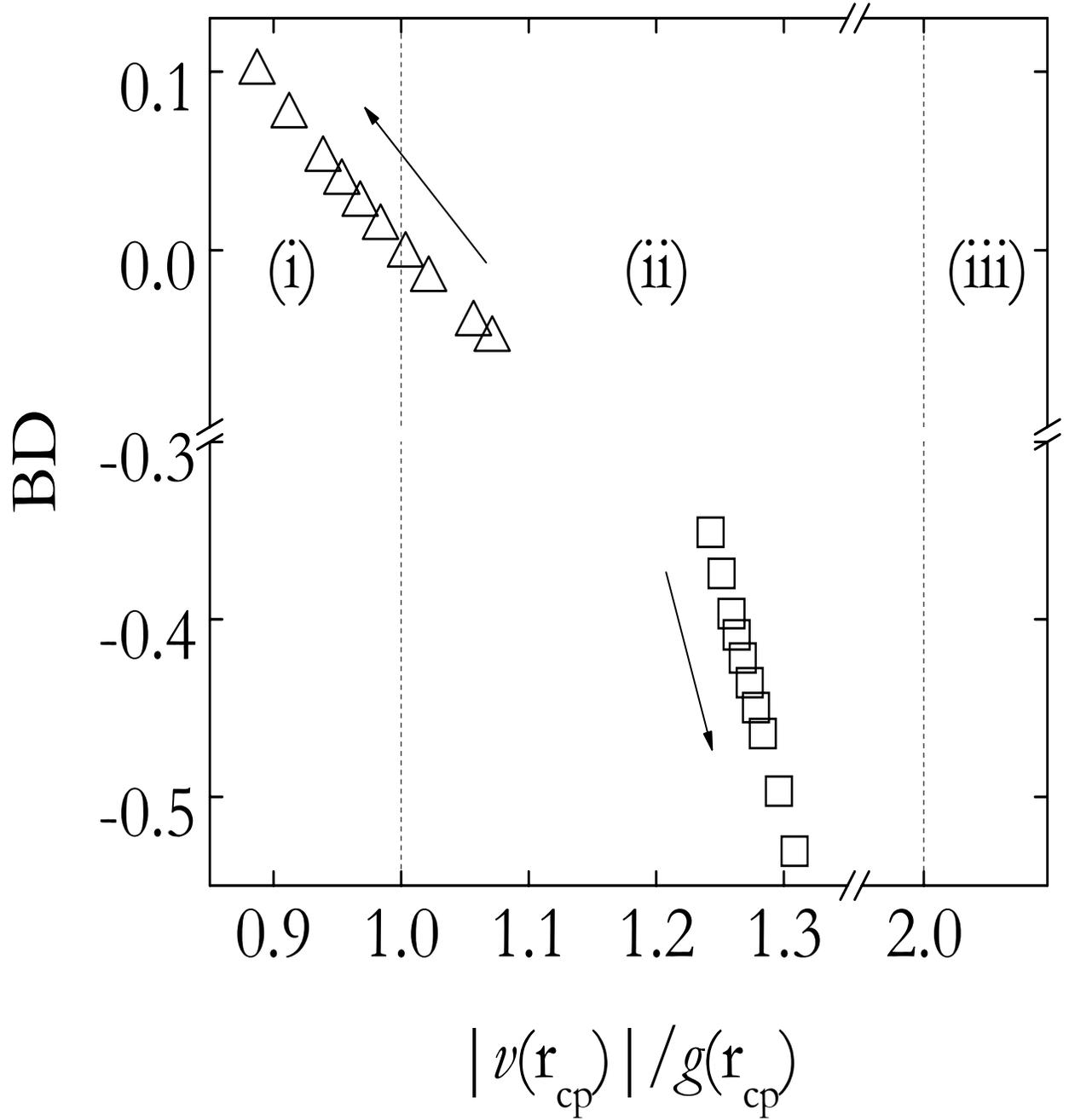}
\caption{The bond degree parameter $BD=h(r_{cp})/\rho(r_{cp})$ vs.
$|v(r_{cp})|/g(r_{cp})$ ratio for the Ti-O (squares) and Pb-O
(triangles) bonds. The arrows indicate the direction of pressure
increase. Vertical dashed lines indicate the regions of stability
for the three classes of atomic interactions (see text).
}\label{fig:BD}
\end{figure}

\begin{figure}
\includegraphics[width=1\textwidth,height=.95\textheight]{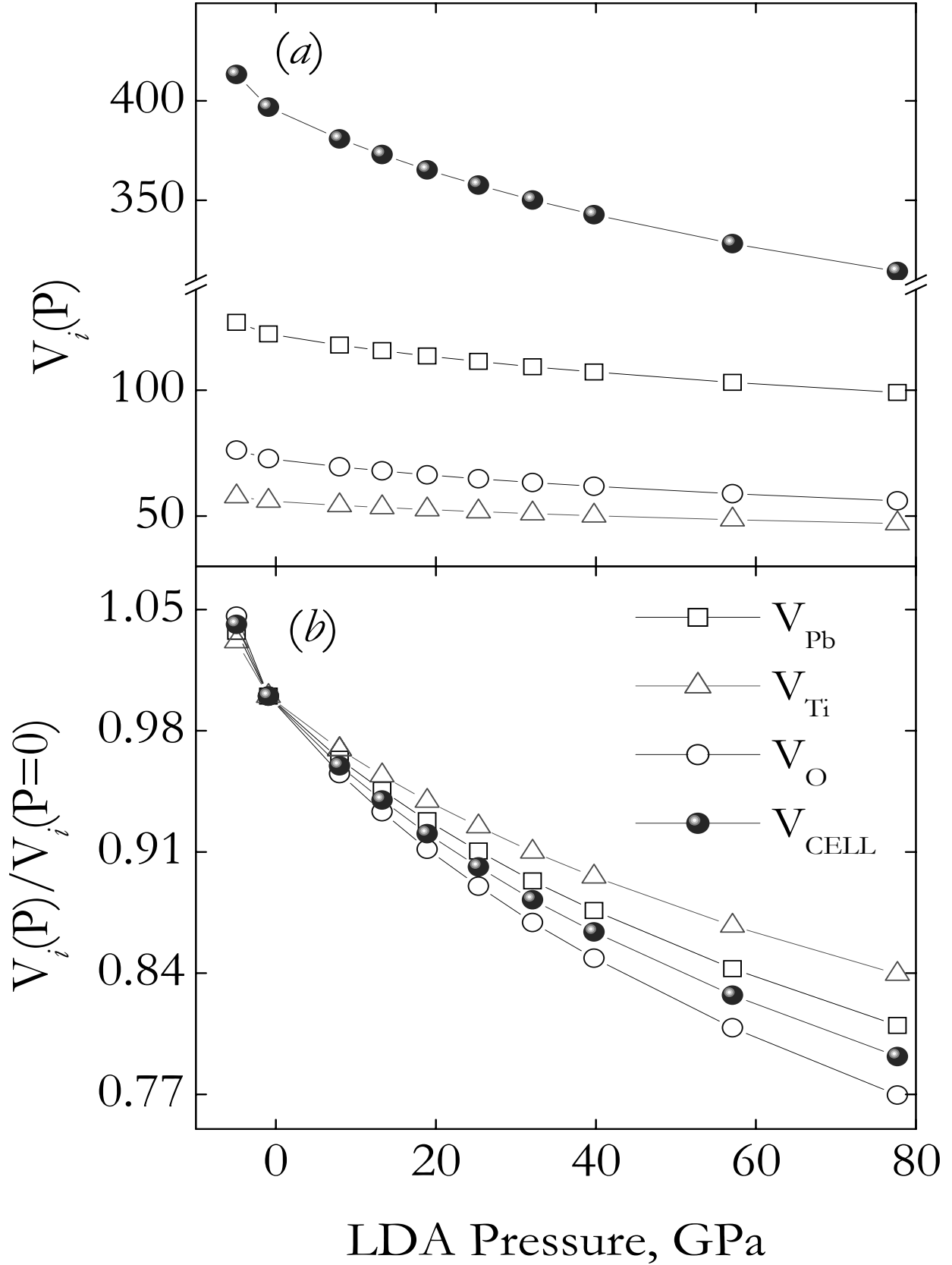}
\caption{(a) Pressure evolution of the ionic Bader volumes for Pb,
Ti and O and the unit cell volume V$_{CELL}$ in the cubic phase of
PbTiO$_3$. (b) The pressure dependence of the normalized volumes.
(The normalized volumes are equal to unity at 0 GPa.)
}\label{fig:Vol}
\end{figure}

\begin{figure}
\includegraphics[width=1\textwidth,height=.75\textheight]{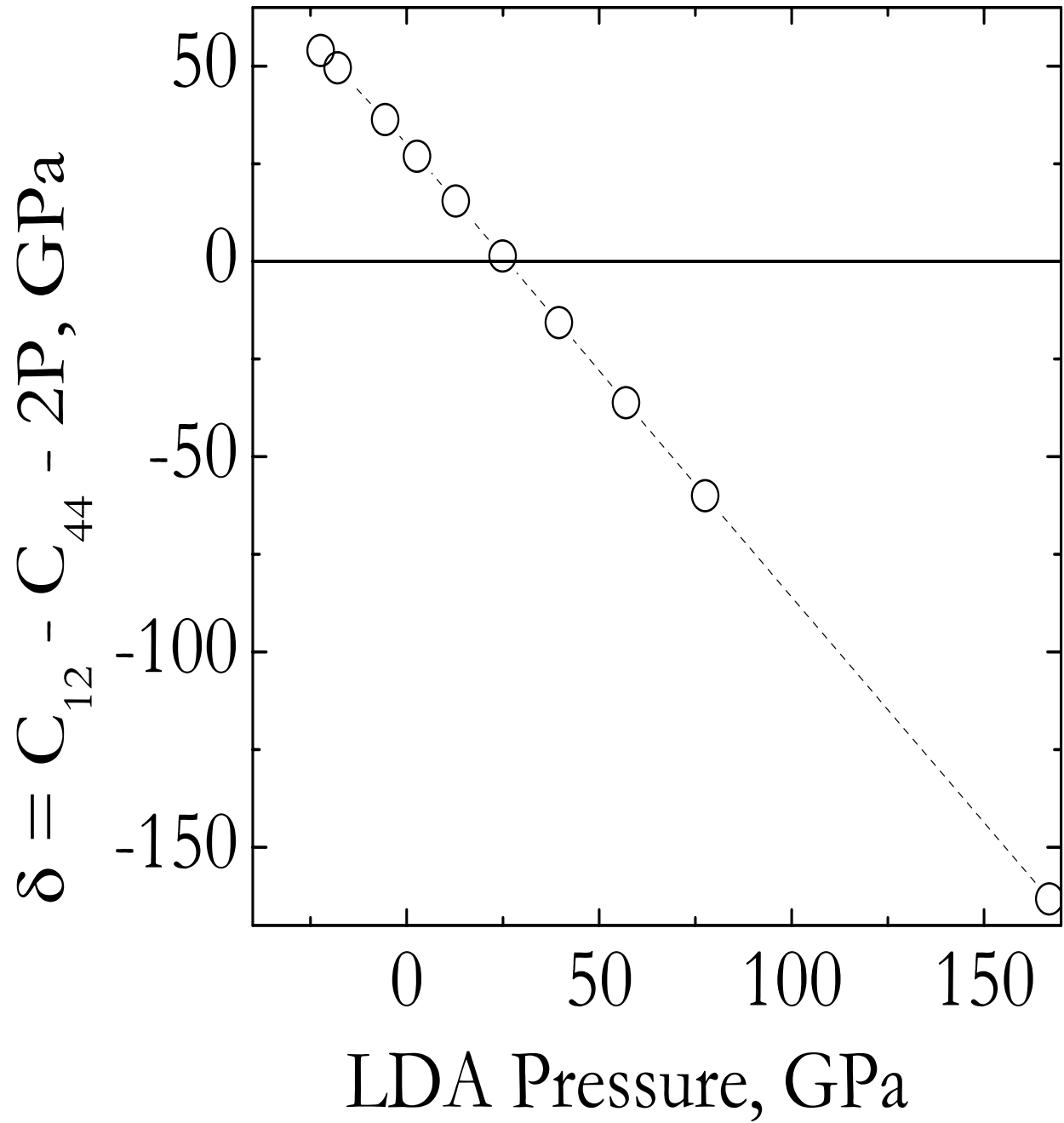}
\caption{LDA-predicted deviations from the Cauchy relation
($\delta = C_{12} - C_{44} - 2P)$ as a function of pressure for
the cubic phase of PbTiO$_3$.} \label{fig:Cauchy}
\end{figure}

\begin{figure}
\includegraphics[width=1\textwidth,height=.75\textheight]{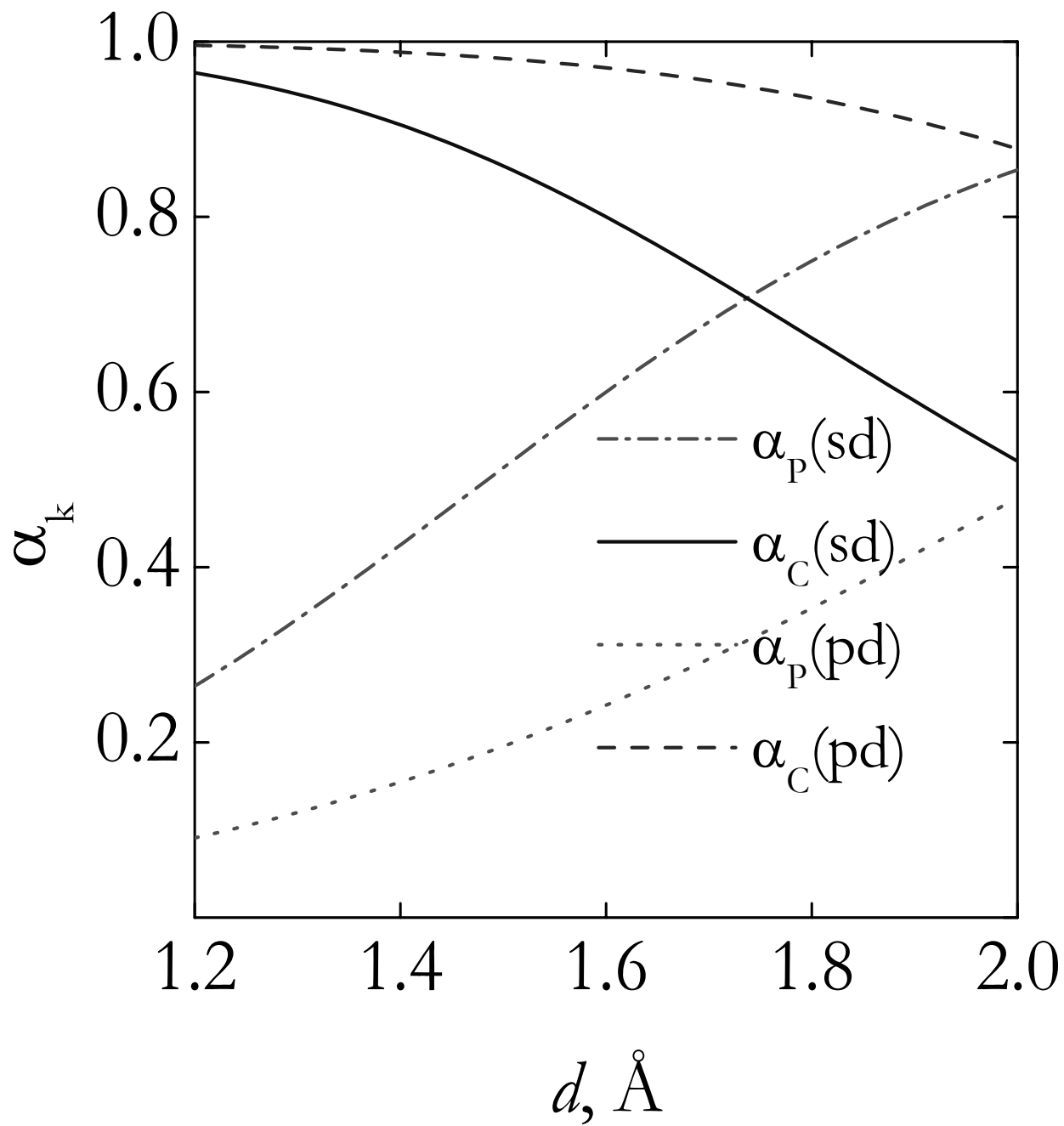}
\caption{The Ti$3d$-O$2s$ and Ti$3d$-O$2p$ bonds polarity and
covalency of a perovskite as a function of the interatomic Ti-O
distance $d$ in the ideal cubic phase. (Note that the pressure
increases as $d$ decreases).}\label{fig:Alphas}
\end{figure}

\begin{figure}
\includegraphics[width=1\textwidth,height=.75\textheight]{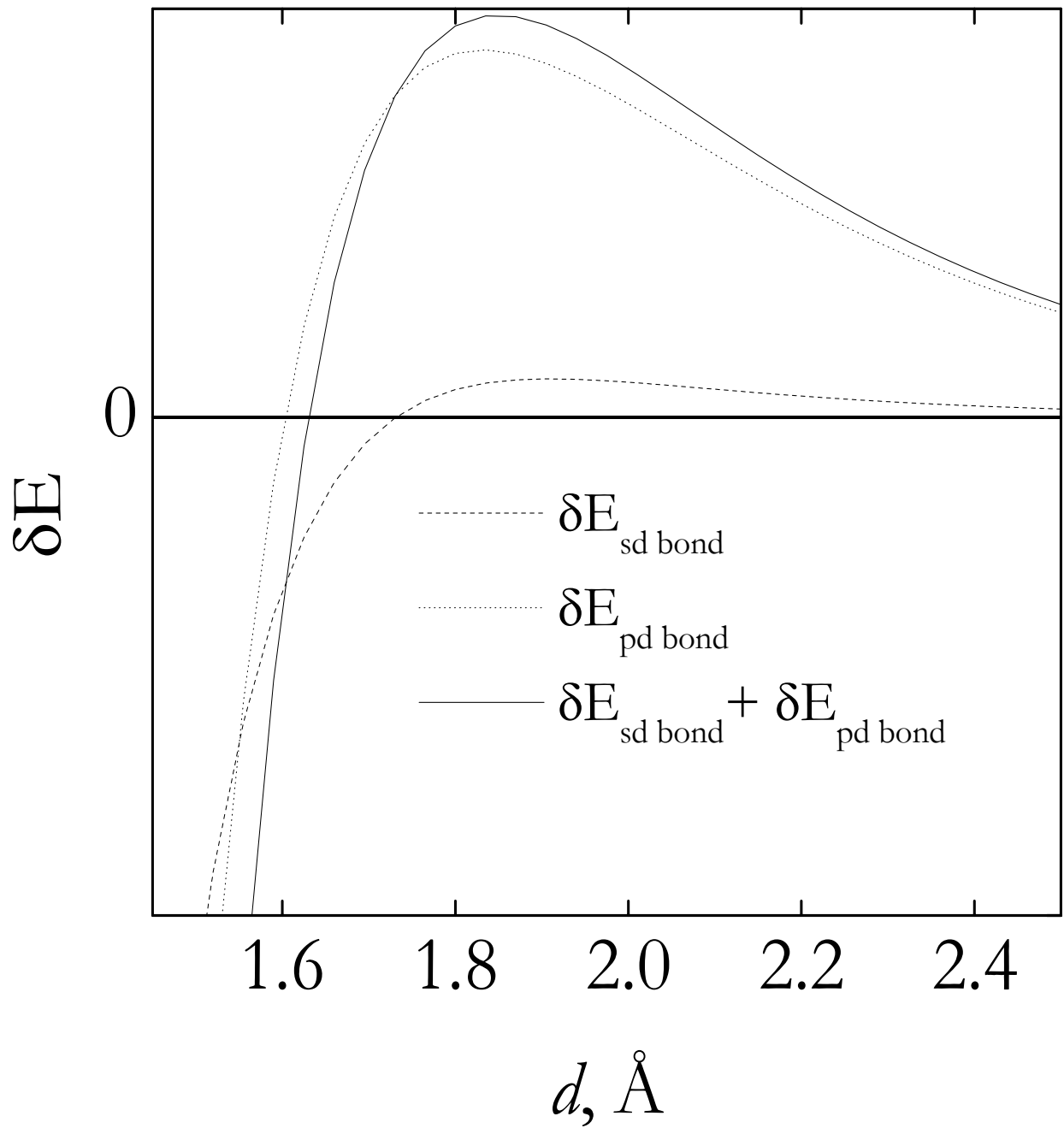}
\caption{Change in bonding energy (solid lines) within the
bond-orbital model of Harrison. This change is directly
proportional to the soft-mode frequency square within this model.
The contributions of the $sd$ and $pd$ bonds to that total change
are also indicated via dashed lines. $d$ is the Ti-O distance in
the ideal cubic phase.}\label{fig:Froz}
\end{figure}

\begin{figure}
\includegraphics[width=1\textwidth,height=.95\textheight]{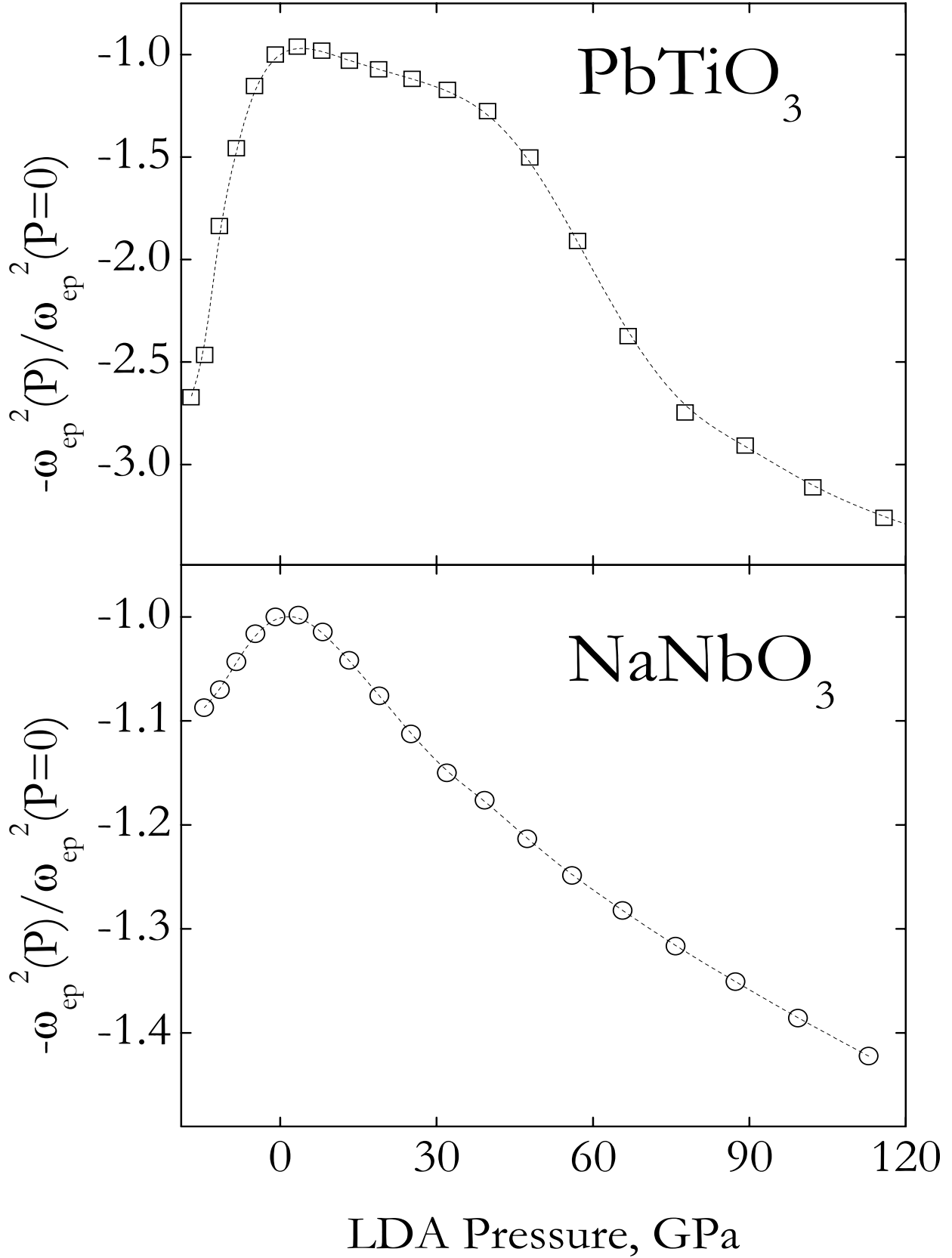}
\caption{Normalized electron-phonon contribution $\omega^2_{ep}$
to the ferroelectric soft mode at the $\Gamma$ point in the cubic
state of PbTiO$_3$ and NaNbO$_3$. Normalization has been done such
as to obtain values of $-1$ at 0 GPa.} \label{fig:D2}
\end{figure}

\end{document}